\documentclass[sigconf]{acmart}

\usepackage{graphicx}
\usepackage{textcomp}
\usepackage{xcolor}
\usepackage{subcaption}
\usepackage{svg}
\usepackage{float}
\usepackage{multicol}
\usepackage{array}
\usepackage{mwe}
\usepackage[noend]{algpseudocode}
\usepackage[ruled, lined, longend]{algorithm2e}
\usepackage{multirow}
\usepackage{arydshln}
\usepackage{adjustbox}
\usepackage[thinc]{esdiff}
\usepackage{booktabs} 
\usepackage{pifont}
\usepackage{hyperref}
\usepackage{todonotes}
\usepackage{enumitem}
\usepackage{tablefootnote}
\usepackage{threeparttable}
\usepackage{booktabs}
\usepackage{multirow}
\usepackage{amsmath}
\usepackage{siunitx}
\usepackage{graphicx}

\AtBeginDocument{%
  }

\setcopyright{none}
\acmConference[]{}

\begin{document}




\title{TeLLMe v2: An Efficient End-to-End \underline{Te}rnary \underline{LLM} Prefill and Decode Accelerator with Table-Lookup Matmul on \underline{E}dge FPGAs}
\author{Ye Qiao}
\authornote{Equal contribution.}
\email{yeq6@uci.edu}
\affiliation{%
  \institution{University of California, Irvine}
  \city{Irvine}
  \state{California}
  \country{USA}
}

\author{Zhiheng Chen}
\authornotemark[1] 
\email{zhihenc5@uci.edu}
\affiliation{%
  \institution{University of California, Irvine}
  \city{Irvine}
  \state{California}
  \country{USA}
}

\author{Yifan Zhang}
\email{yifanz58@uci.edu}
\affiliation{%
  \institution{University of California, Irvine}
  \city{Irvine}
  \state{California}
  \country{USA}
}

\author{Yian Wang}
\email{yianw11@uci.edu}
\affiliation{%
  \institution{University of California, Irvine}
  \city{Irvine}
  \state{California}
  \country{USA}
}

\author{Sitao Huang}
\email{sitaoh@uci.edu}
\affiliation{%
  \institution{University of California, Irvine}
  \city{Irvine}
  \state{California}
  \country{USA}
}

\begin{abstract}
With the emergence of wearable devices and other embedded systems, 
deploying large language models (LLMs) on edge platforms becomes an urgent need. However, it is challenging because of their high computational and memory demands. Although recent low-bit quantization methods (e.g., BitNet, DeepSeek) compress weights to as low as 1.58 bits with minimal accuracy loss, edge deployment is still constrained by limited on-chip resources, power budgets, and the often-neglected long latency of the prefill stage. 
We present TeLLMe, the first table-lookup-based ternary LLM accelerator for low-power edge FPGAs that fully supports both prefill and autoregressive decoding using 1.58-bit weights and 8-bit activations. TeLLMe incorporates our proposed novel techniques including (1) a table-lookup-based ternary matrix multiplication (TLMM) engine utilizing grouped activations and online precomputation for low resource utilization and high throughput; (2) a fine-grained analysis and optimization onan analytic and fine-grained URAM-based weight buffer management of weight loading from global memory and compute engine weight access; (3) a streaming dataflow architecture that fuses floating-point element-wise operations with linear computations to hide latency; (4) a reversed-reordered prefill stage attention with fused attention operation for high memory efficiency; and (5) a resource-efficient specialized decoding stage attention. Under a 5W power budget, TeLLMe delivers up to 25 tokens/s decoding throughput and 0.45s to 0.96s Time-to-First-Token (TTFT) for 64–128 token prompts, marking a significant energy-efficiency advancement in LLM inference on edge FPGAs. 
\end{abstract}

\maketitle

\section{Introduction}
Large language models (LLMs) have been evolving rapidly, delivering state-of-the-art results in machine translation, code generation, question answering, and conversational AI. Models such as GPT-3~\cite{brown2020language}, LLaMA~\cite{touvron2023LLaMA}, and DeepSeek-R1~\cite{guo2025deepseek} highlight the benefits of scale, but those gains arrive with steep cost increases in computation, memory, and energy.

Deploying LLMs on edge devices (embedded CPUs, GPUs, FPGAs, etc.) preserves privacy, reduces latency, and enables autonomy, yet it is difficult due to tight limits on memory bandwidth and capacity, compute resources, and power budgets. Autoregressive decoding further stresses these constraints: growing key–value (KV) caches, long-context handling, and strict latency requirements often become the dominating performance bottlenecks. 

Extreme model compression via low-bit quantization has emerged as a promising technique~\cite{qiao2022two,10025006}. BitNet~\cite{bitnet,qiao2025cobra, qiao2025tellme} showed that Transformers can be trained with 1-bit weights; BitNet-1.58~\cite{bitnet158} and DeepSeek~\cite{deepseek} extend this idea to ternary quantization (${-1,0,+1}$), approaching full-precision quality while drastically reducing model size and energy. However, closing the gap between algorithmic compression and efficient, end-to-end deployment on actual edge hardware requires a co-design solution that simultaneously optimizes compute, memory, and scheduling.

While researchers have proposed several solutions to accelerate the decoding stage of LLMs, the existing works often ignore the significant  latency of the prefill stage, making prefill a critical performance bottleneck in the end-to-end LLM systems. 
For example,~\cite{li2025pushing} demonstrates efficient decoding on embedded FPGAs but leaves the prefill stage largely unaddressed. On device, prefill latency directly impacts users' perceived responsiveness and safety; it is not hidden behind cloud-scale parallelism. 
Therefore, prefill should also be treated as a first-class citizen and carefully accelerated  alongside decoding when designing an edge accelerator. 


We present \textbf{TeLLMe}—the \textbf{Te}rnary \textbf{L}arge \textbf{L}anguage \textbf{M}odel \textbf{e}dge accelerator—an edge FPGA accelerator, to our knowledge, the first to incorporate a table-lookup (TL) matrix multiplication (MatMul) approach tailored for ternary LLM inference with full support for both \emph{prefill} stage and \emph{decoding} stage. TeLLMe targets cost-effective low-power FPGAs (AMD Kria KV260), supports 1.58-bit (ternary) weights and 8-bit activations, and co-optimizes compute, memory, and scheduling for low-latency, energy-efficient LLM inference.

Prior FPGA accelerators~\cite{LUTNET, SUMLUTNET, TLMAC} leveraged table-lookup techniques to speed up low-precision arithmetics in other domains, demonstrating the viability of lookup table (LUT)-centric accelerators. However, to our knowledge, TeLLMe is the first to apply a \emph{ternary, table-lookup matmul} design to end-to-end LLM inference (prefill and decoding). Furthermore, it integrates LLM-specific optimizations—fused attention for prefill, disaggregated prefill/decoding dataflows, streaming fusion of dequantization, quantization, element-wise operations, etc., and URAM-aware weight orchestration. With these optimizations, our proposed TeLLMe delivers state-of-the-art efficiency on actual FPGA boards.
Our key contributions include:
\begin{itemize}[leftmargin=*, nosep]
\item \textbf{End-to-end accelerator for ternary LLMs.} We build, to our knowledge, the first edge FPGA accelerator that employs a table-lookup matmul engine for ternary LLMs with full support for both prefill and decoding, achieving up to 25 tokens/s generation throughput and up to 143 tokens/s prefill throughput while consuming under 5~W.
\item \textbf{Ternary table-lookup matmul.} We propose a table-lookup-based matrix multiplication (TLMM) unit optimized for FPGAs that reuses grouped activations and performs online accumulation for ternary multiplications across attention projections and the feed-forward network (FFN).
\item \textbf{URAM-aware weight orchestration.} We introduce fine-grained URAM buffering for high-throughput weight streaming from off-chip memory and provide an analytic method to optimize TLMM engine parameter selection under URAM and LUT constraints.
\item \textbf{Streaming fusion with mixed precision.} We fuse floating-point (FP) dequantization, integer (INT) quantization, and elementwise operations (residual add, rotary embedding, etc)  around the INT-based TLMM to enable mixed-precision execution and overlap compute with dataflow architecture.
\item \textbf{Disaggregated prefill and decoding attention.} We separate the attention pipelines to match their distinct compute/memory patterns. For prefill, we design a fused attention unit with a reversed-attention mechanism and a fully fused pipeline that minimizes off-chip traffic and avoids redundant masked computation. For decoding, we exploit on-chip memory to retain intermediate softmax scores and prevent DDR reloads.
\end{itemize}
In summary, TeLLMe demonstrates that a LUT/URAM-native, table-lookup matmul engine co-designed with LLM-specific dataflows can unlock end-to-end ternary inference on resource-constrained FPGAs. To our knowledge, it is the first accelerator to realize table-lookup matmul for LLMs with comprehensive support for prefill and decoding on real hardware, establishing a strong baseline for energy-efficient, low-latency generative AI at the edge.

\section{Background and Related Work}

\subsection{Transformer/LLM Acceleration on Edge}
Deploying Transformers on FPGAs is challenging due to limited bandwidth and logic resources. Several works have explored quantized Transformer accelerators on embedded FPGAs/CPUs. 
\textbf{T-MAC} \cite{tmac} implements a TLMM kernel for CPUs using low-bit weights and high-bit activations. It achieves notable performance on edge CPUs. Moreover, the prefill speed is not ideal and close to the decoding throughput, suffering from the lack of parallelism computation bound inside edge CPUs. \textbf{{Li et al.}}~\cite{li2025pushing} successfully implemented a 4-bit quantized LLaMA2-7B model on the AMD KV260 platform. Although the model weights are quantized to 4-bit, the decoding computations rely on unquantized FP16 activations, thereby requiring all operations to be conducted in FP16. Moreover, hardware acceleration is limited to the decoding stage and does not address the computational demands of the prefill stage. \textbf{LLaMAF}~~\cite{LLaMAf} targets TinyLLaMA model with int8 quantization on ZCU102 with speed of 1.5 tokens/s. It leverages pipelined matrix-vector units on FPGA and asynchronous scheduling for non-linear operators on ARM A53 cores, but suffers from ARM-FPGA communication and ignore the prefill stage. \textbf{Edge-MoE}~\cite{edgemoe} introduces a memory-efficient MoE vision transformer accelerator using dynamic task-level sparsity. 
A key technique is the specialized reordering to enable the data reuse of attention computation for edge FPGAs when the attention score matrix is dense. \textbf{SECDA}~\cite{haris2024designing} designs the matrix multiplication accelerator supporting block FP quantized operations on PYNQ, reducing latency by 11x compared to dual-core Arm NEON-based CPU execution for the TinyLLaMA model. However, the tokens/s is only 0.58, which means 2 seconds for one token generation.
\textbf{MEADOW}~\cite{moitra2025meadow} presents a memory-efficient weight packing framework for low-power edge LLMs, implemented on the Xilinx ZCU102 FPGA, with approximately 10W power consumption. It introduces a weight packing strategy for memory-efficient linear operation in both prefill and decoding stages. However, they only implement the INT8 matmul without considering the end-to-end acceleration of FP16 smooth and dequant operation that supports the utilized INT8 smoothquant~\cite{xiao2023smoothquant}.  
It reaches the prefill speed up to approximately 0.64 ms at 64 tokens and decoding speed up to 2 token/s for the OPT 1.3B model from the profiling result. Compared to these, our work is the first to unify binary weight inference, high throughput prefill/decoding support, and FPGA memory hierarchy optimization into one cohesive design.



\subsection{LLMs on High-performance FPGAs}
\textbf{LightMamba}~\cite{LightMamba} introduces an FPGA-friendly post-training quantization algorithm for the Mamba2-2.7B model, supporting W4A4 quantization. Deployed on high-end Versal VCK190 and U280 FPGAs, it achieves throughputs of 7.21 and 93 tokens/s, respectively.
\textbf{FlightLLM}~\cite{FlightLLM} deploys the LLaMA2-7B model on Alveo U280 and Versal VHK158 FPGAs, outperforming NVIDIA GPUs in single-batch inference with a throughput of 153 tokens/s.
\textbf{EdgeLLM}~\cite{EdgeLLM} utilizes the VCU128 FPGA with HBM to deploy LLaMA2-7B, achieving a higher decoding speed than FlightLLM.
\textbf{TerEffic}~\cite{yin2025tereffic} deploys a ternary 1.3B non-transformer model on the Alveo U280, achieving a remarkable 1400 tokens/s. However, its reliance on soft logic for ternary matrix multiplication, rather than a table-lookup approach, results in over 700K LUTs consumed for redundant paralleled ternary matmul.
These works leverage high-end FPGAs with over 500K LUTs, thousands of DSP units, thousands of BRAM cells, HBM providing 460 GB/s bandwidth, and power consumption exceeding 50W, with costs exceeding \$10,000 (vs. \$248 for the KV260)~\cite{AMDVersalOverview}. While these platforms maximize FPGA performance, their performance is not directly comparable to edge FPGAs.
\subsection{Existing Table Lookup Works on FPGAs}
Using DSP48E1 blocks for 1.58-bit TLMM on FPGAs is highly inefficient due to their design for high-throughput, fixed-point arithmetic (25x18-bit multiplier, 48-bit accumulator). Ternary operations, with weights of -1, 0, or 1, require only simple selection or negation, underutilizing the DSP’s complex logic. Instead, LUTs can efficiently handle “pass,” “negate,” or “zero” operations, making DSPs a resource- and power-inefficient choice for highly quantized neural networks better suited to FPGA fabric logic.
At a larger architectural scale, implementing ternary matmul with purely selection-based logic also becomes inefficient in terms of resource utilization and latency. While ternary weights simplify multiplication into selections, achieving high parallelism demands numerous LUT-based multiplexers. This approach is suboptimal for large models (i.e., with large $d_{\text{model}}$), where the limited set of ternary operations results in repetitive computational patterns \cite{TLMAC}. 
As an alternative to logic-based acceleration, a table-lookup method has been proposed as an efficient solution for ternary matmul on CPUs ~\cite{tmac}. This technique leverages specialized SIMD instructions, such as ARM NEON and AVX, for rapid parallel look-ups. However, its effectiveness is often constrained by the limited size of the look-up table (256 bit/128 bit), which can lead to frequent, high-latency memory accesses when the table is swapped.

FPGAs provide an optimal solution by leveraging their inherent LUT resources to implement larger tables using LUT-based Distributed RAM. As demonstrated in studies such as~\cite{LUTNET, LOGICNET,SUMLUTNET}, each basic LUT can be configured to store values that facilitate neural network inference. However, fine-grained control of LUT resources is currently limited to simple tasks like MNIST and smaller network sizes in the thousands level of neurons. To expand the area of TL-based computing on FPGA,~\cite{TLMAC} introduces a CNN-specific TL engine that adopts a coarse-grained approach to distributed LUTRAM. Yet, this approach focuses solely on TL convolution logic, overlooking comprehensive dataflow considerations for real-world INT/FP mixed-precision quantized neural network workloads. Moreover, they lack the design in TL logics' interactions with on-chip and off-chip smemory, which is vital in the memory bound trait of transformer and LLM inference. To the domain to the LLM era, we propose a TLMM engine for LLM inference on FPGAs, integrated with a detailed exploration of efficient mix-precision design and fine-grained on-chip URAM management. 

\begin{figure}[t] 
\centering
\includegraphics[width=\linewidth]{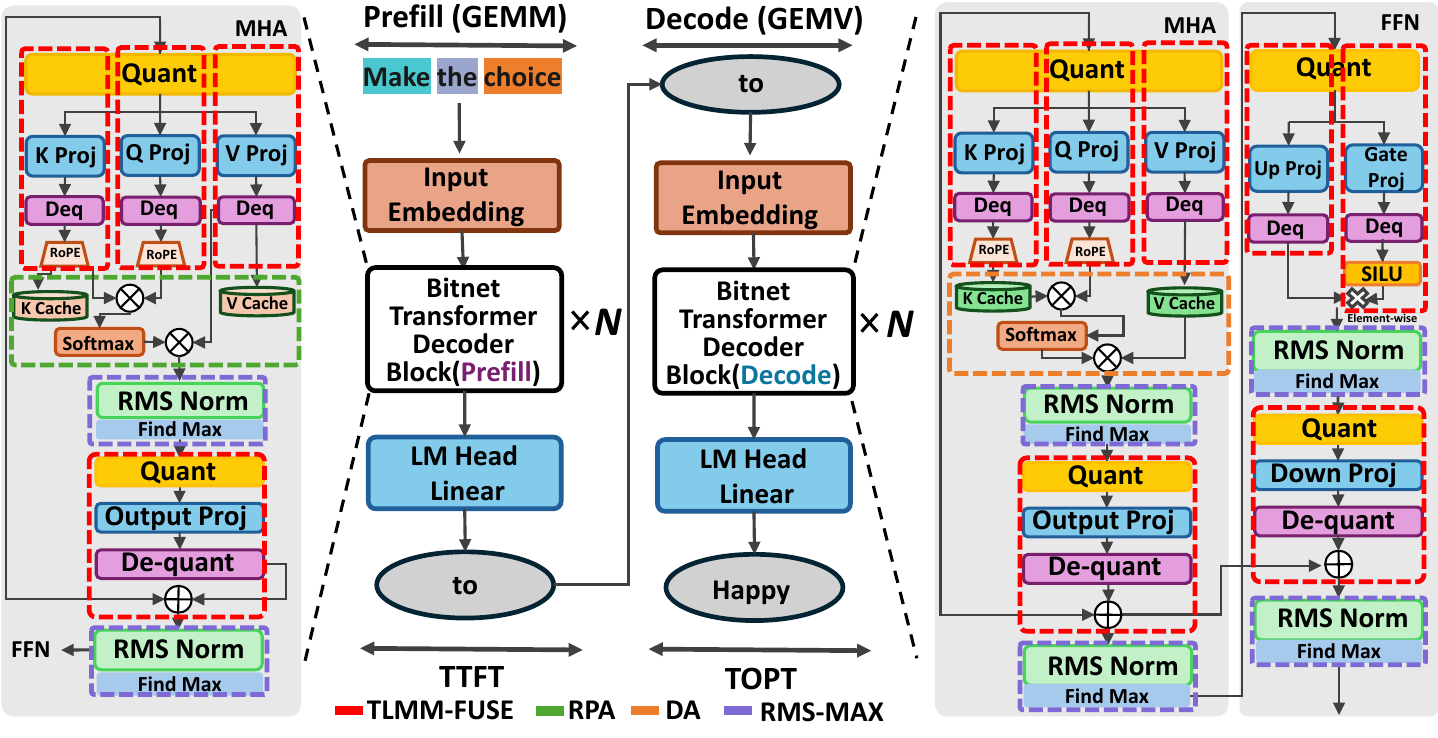}
\caption{Breakdown of TeLLMe 1.58-bit Model Inference Process with Prefill and Generation}
\label{Fig:Bitnet}
\end{figure}

\section{TeLLMe Design Methodology}

\begin{figure}[t] 
\centering
\includegraphics[width=\linewidth]{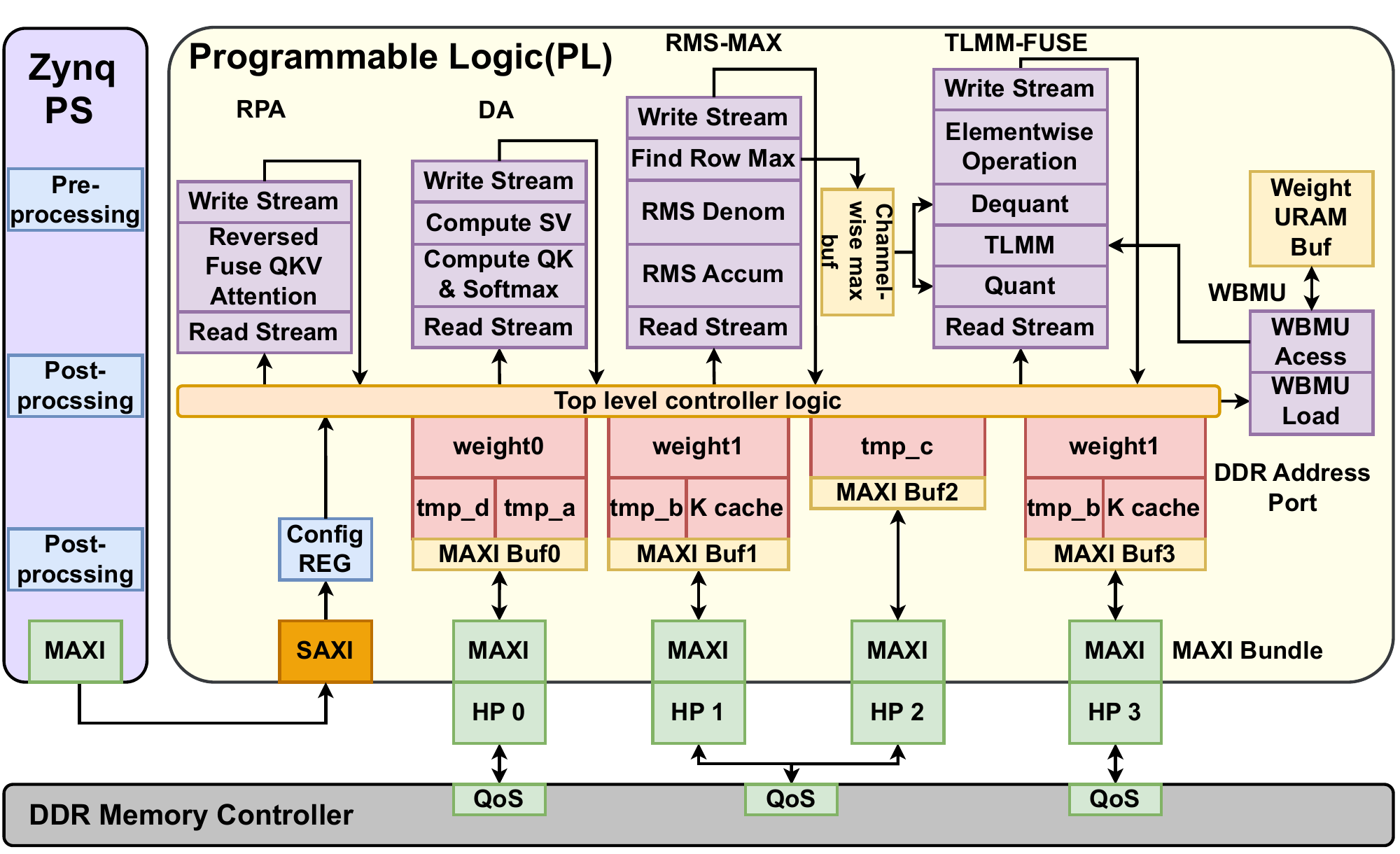}
\caption{System Architecture of TeLLMe}
\label{fig:sys_arch} 
\end{figure}

\subsection{Overview}
The technical challenges of ternary LLM implementation on edge FPGA are shown as follows, 
\textbf{\ding{182} Challenge of implementing TLMM for Ternary LLM on FPGA:} The TLMM approach is well-suited for implementing ternary linear operations. However, the current TL-based method on FPGA~\cite{TLMAC, LOGICNET} focuses solely on TL-based convolution or small-sized perception machine, overlooking its interaction with on-chip and off-chip memory. This interaction is critical for LLMs, which are memory-bound and heavily relying on efficient memory management.
\textbf{\ding{183} Hardware Support for Mixed-Precision in Low-bit Quantization:} Modern quantization strategies often employ a hybrid-precision approach, where weights and activations are represented with both FP16 (attention, dequant, quant, rotary embedding, etc) and quantized low-bit integers (linear). 
Designing a unified hardware architecture that can process these varied data types without incurring significant performance or area overhead is a complex task.
\textbf{\ding{184} Diverse Computational Patterns in Attention Mechanisms:} The attention mechanism in transformer models exhibits two different computational profiles. The initial \textbf{prefill} phase is compute-bound and characterized by parallel processing of the input prompt, while the subsequent \textbf{decoding} phase is memory-bound and involves sequential, auto-regressive token generation. A successful FPGA accelerator should efficiently handle both distinct patterns.


The overall Bitnet model architecture is depicted in Fig. \ref{Fig:Bitnet}, comprising attention computation, linear projections with quantization and dequantization, rotary position embedding (RoPE), RMSNorm, absolute maximum (ABSMAX) quantization, and SwiGLU modules. During the prefill phase, the KV projections are stored in the KV cache, whereas in the decoding phase, the cached KV values are loaded for attention computation.
As illustrated in Fig. \ref{fig:sys_arch}, the accelerator architecture comprises the following key modules to address the challenges: (1) a TLMM engine that handles both decoding and prefill passes, integrated with element-wise operations and quantization, referred to as the TLMM-FUSE unit to address \ding{182}, \ding{183}; (2) an on-chip weight buffer management unit (WBMU) that handles the off-chip DDR weight data transfer and accessing of the URAM weight buffer to address \ding{182}; (3) a reversed-reordered prefill attention (RPA) unit that get rid of the redundant attention mask during prefill stage to address \ding{183}, \ding{184}; (4) a decoding attention (DA) unit that match the vector-wise computation in decoding stage to address \ding{183}, \ding{184}; and (5) an RMSNorm unit combined with a find-max operation for supporting the ABSMAX quantization, refered to as RMS-MAX unit to address \ding{183}. 

All modules are designed in a dataflow style to enable function-level pipelining. The DDR address ports are bound to distinct MAXI bundles, facilitating independent data streaming from DDR memory. Four temporary address ports are allocated for intermediate storage in DDR. The weight address ports handle weight loading, while the K cache and V cache address ports manage the prefill attention KV return values and decoding attention input values, respectively. The relationship between these modules and the model's software logic is illustrated in Fig.~\ref{Fig:Bitnet}.

\subsection{Table-Lookup Matmul (TLMM) Engine}

\subsubsection{TLMM Engine Design} \label{sec:TLMM design}
\begin{figure}[h] 
\centering
\includegraphics[width=\linewidth]{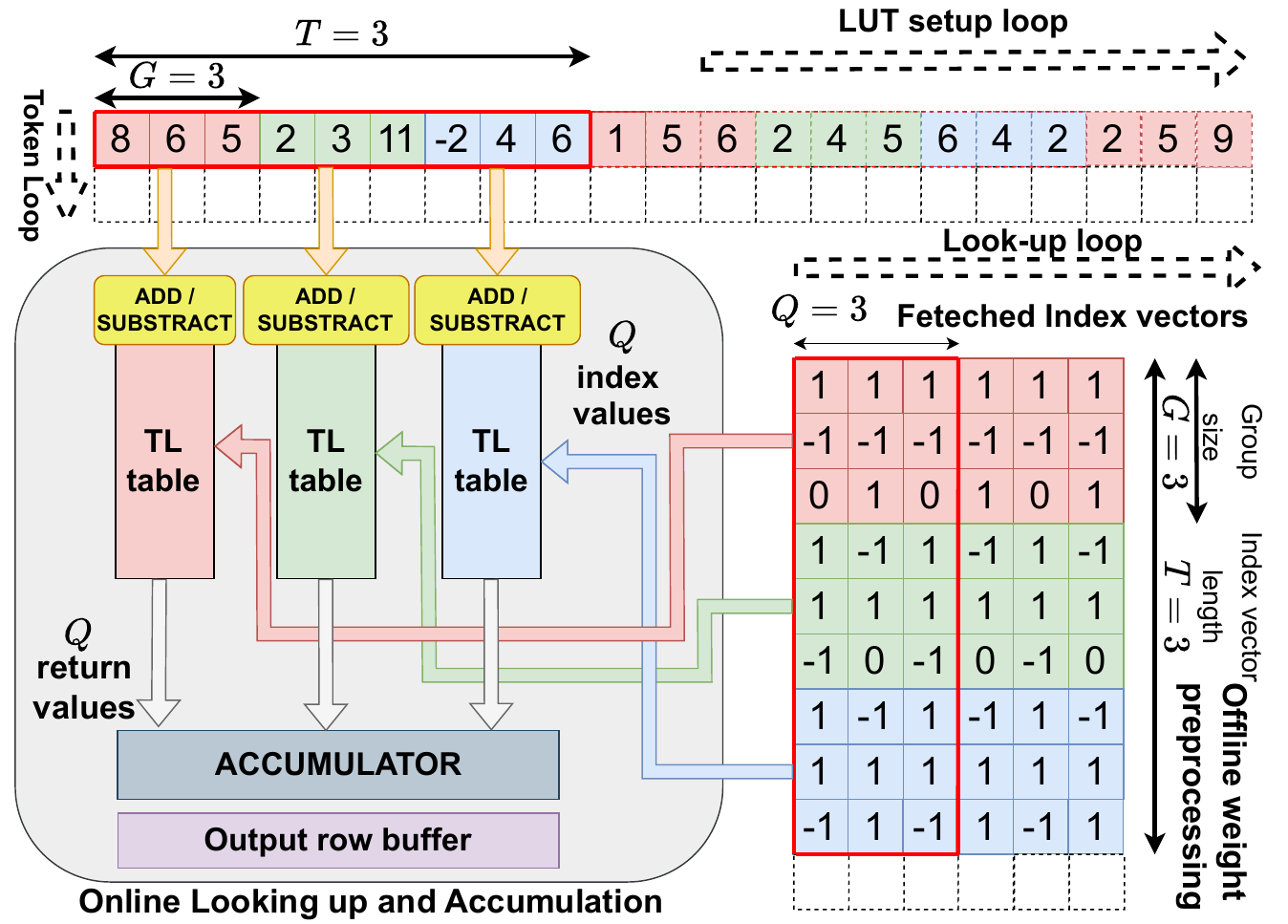}
\caption{High Level Dataflow of TLMM ($G = 3, T=3,Q=3$)}
\label{fig:lutmatrix multiplication} 
\end{figure}

The TLMM method, illustrated in Fig. \ref{fig:lutmatrix multiplication}, consists of two main stages: offline weight preprocessing and online matrix multiplication. In the offline stage, the ternary weight matrix \({\bf W} \in \{-1, 0, 1\}^{n \times k}\) is partitioned into groups of size $G$, which are then encoded into compact indices. This encoding yields \(N_{\text{TB}} = 3^{G}\) unique combinations per group, requiring an index bitwidth of \( B_{\text{idx}} = \lceil \log_2(3^G) \rceil \) bits.

During the online stage, the matrix-vector product between the activation matrix \({\bf A} \in \text{INT8}^{m \times n}\) and the preprocessed weight indices is computed. For each output vector $\bf o$, a group of $G$ INT8 activation values is fetched to dynamically generate a small TL table implemented by distributed RAM. This table is populated by a pre-computation unit of adder/subtractor trees and stores all \(N_{\text{TB}}\) possible partial sums for the current activations, with each entry sized at a bitwidth of \(B_{\text{TB}} = 8 + \lceil \log_2 G \rceil\) to prevent overflow. The preprocessed weight indices are then used to look up these partial sums from the TL table. To support the parallel multiple read requests of $Q$ of $T$ TL tables, each TL table has to duplicate $Q$ times, resulting in the TL table size of $B_{\text{TB}} \times N_{\text{TB}}\times Q \times T$. Finally, these values are accumulated to produce the output vector \({\bf o} \in \text{INT32}^{1 \times k}\), which is pushed into the stream first-in-first-out unit (FIFO) channel for subsequent operations. 
To better vectorize the TLMM and facilitate efficient on-chip URAM access, the consecutive \( T \) indices can be grouped into an index vector, enabling simultaneous access to different TL tables. The weight index vector matrix can be rewritten as \( {\bf W}_{\text{idx}} \in \{B_{\text{idx}}, T\}^{\frac{n}{T\times G} \times k} \).

\begin{figure}[t]
    \centering
    \begin{subfigure}[b]{0.47\textwidth}
        \centering
        \includegraphics[width=\linewidth]{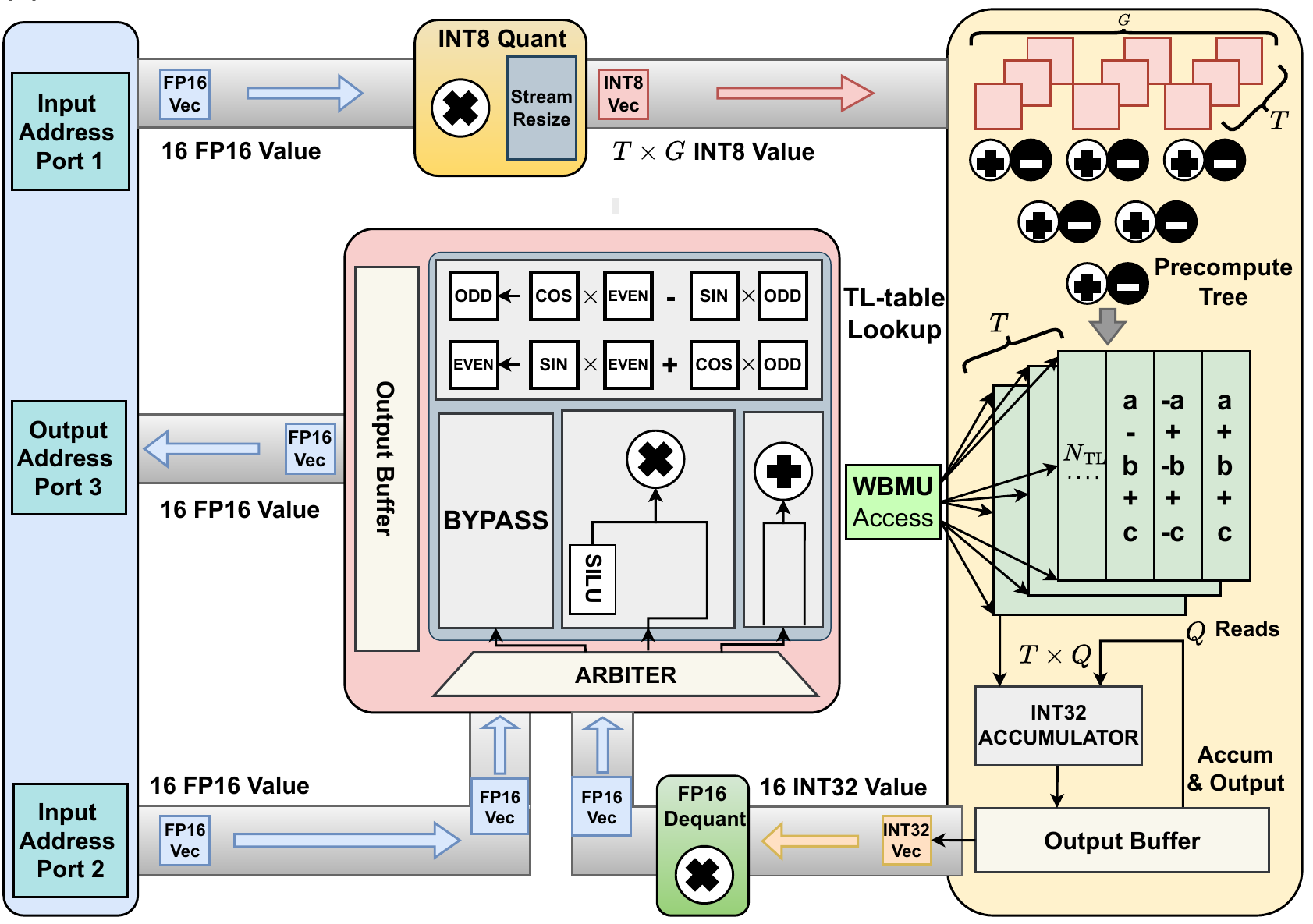}
        \vspace{-2mm}
        \caption{Detailed Design of TLMM-FUSE Unit}
        \label{fig:TLMM_FUSE}
    \end{subfigure}
    \hfill 
    \begin{subfigure}[b]{0.27\textwidth}
        \centering
        \includegraphics[width=\linewidth]{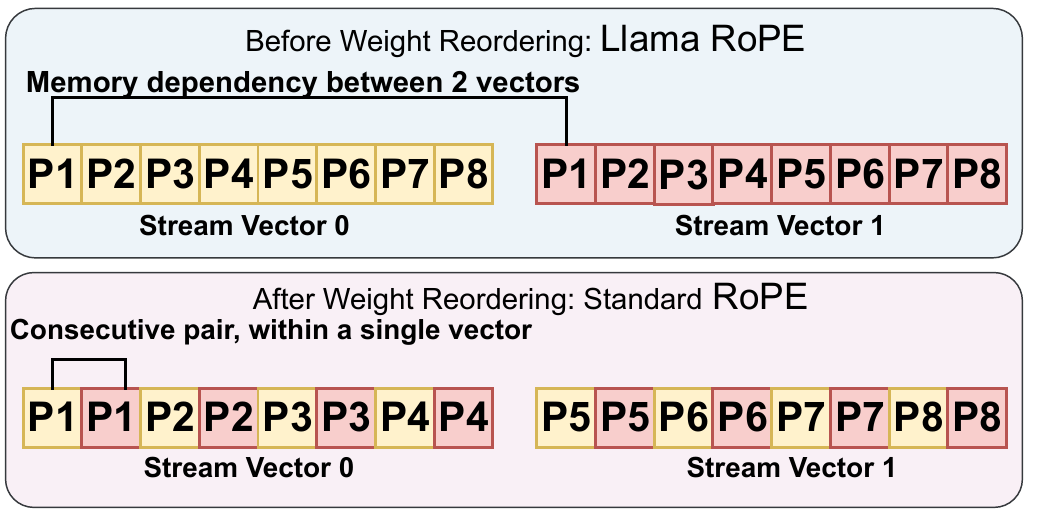}
        \vspace{-2mm}
        \caption{Mem. Dependency of Different RoPE} 
        \label{fig:Rope}
    \end{subfigure}
    \begin{subfigure}[b]{0.20\textwidth}
        \centering
        \includegraphics[width=\linewidth]{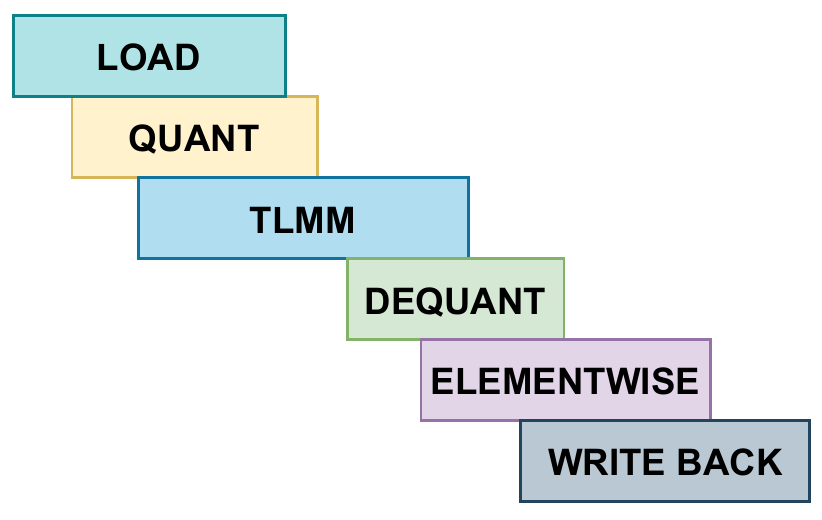}
        \vspace{-2mm}
        \caption{Dataflow Overlapping}
        \label{fig:TLMM_schedule}
    \end{subfigure}
    \vspace{-2mm}
    \caption{Overview of TLMM hardware components and dataflow. (a) The TLMM-FUSE unit design. (b) The comparison of different RoPE operations. (c) The dataflow execution schedule.}
    \label{fig:combined_designs}
\end{figure}

Regarding the scheduling in Fig. \ref{fig:lutmatrix multiplication} with $G=4$, $T=3$, and $Q=3$ and the detailed architecture in Fig. \ref{fig:TLMM_FUSE}, the innermost loop first performs vector operations to establish the \( T \) TL tables through the precompute tree in parallel, based on the first unpacked \( T \times G \) entries of the \(\bf A \) matrix from the stream FIFO channel. Then, leveraging the multiple reading traits of multiple independent partitioned URAM blocks (array partition strategy and parameter will be detailed in the \ref{SEC:WBMU}), \( Q \) \( {\bf w}_{\text{idx}} \in {\bf W}_{\text{idx}} \) index vectors are processed in parallel for TL table addressing, returning \( Q \times T \) outcomes. The corresponding TL table return values are then accumulated in INT32 into an output buffer of size \( k \). The \( k \) index vectors on each row of \( W \) are traversed in steps of \( Q \). The TL table addressing and accumulation process can be fully pipelined with an interval of one cycle, as there are no inter-iteration dependencies. After the first \( T \times G \) entries of the \( \bf A \) matrix are processed, the \( m \) values of the row of \( \bf A \) are traversed in steps of \( T \times G \) in the intermediate loop. Finally, the outermost loop traverses the different channels in \(\bf A \), corresponding to the tokens in the prefill stage of the LLM. 


 As for the interaction with the weight buffer, \( {\bf W}_{\text{idx}}\) is loaded onto the on-chip URAM in a single pack of DDR loading requests for each ternary linear operation and fully decouple with the computation of TLMM to prevent the compute engine from stalling. Moreover, our TLMM engine supports 3 sizes in the TLMM: $q,k,v,o$ projection sized $d_{\text{model}} \times d_{\text{model}}$, up and gate projection in FFN sized $d_{\text{model}} \times d_{\text{ffn}}$, and down projection in FFN sized $d_{\text{ffn}} \times d_{\text{model}}$. This is also supported by the WBMU to convert the high-level address in weight matrices into a physical URAM access address, and additional zero-padding weight indices will be added to support the $T \times G$ parallelism and transpose relationship between up and down projection. The detailed architecture will be presented in Sec. \ref{SEC:WBMU}.

\subsubsection{Comparison between Different Ternary Matrix Multiplication Designs} \label{sec:TLMM_METHODS}
In general, three methods can be employed to implement TLMM using FPGA fabric.
Method 1 is a naive implementation that employs straightforward selection logic to pass, negate, or zero out the input value based on the ternary weight.
Method 2 partially stores results in a TL table. By exploiting the symmetry of ternary weights (+1, –1), only entries corresponding to positive weights are stored. The number of required entries is given by \( N_{\text{TB}} = (3^G - 1)/2 \), where \( G \) is the group size and the subtracted unit accounts for the all-zero state. However, to support \( T \times Q \) parallel reads, this scheme requires extra logic to identify negative indices and invert the corresponding outputs retrieved from the table.
The third method—our implementation—stores all possible entries in the TL table, totaling \( N_{\text{TB}} = 3^G \). This eliminates the need for additional selection logic and enables a purely lookup-based addressing scheme for \( T \times Q \) parallel reads.
As for resource analysis, the LUT resources for precomputation can be estimated as  
\begin{align}
\text{LUT}_{\text{PRE}} = T \times N_{\text{TB}} \times \text{LUT}_{\text{tree}},
\end{align}
where the $\text{LUT}_{\text{tree}}$ is the average LUT utilized in each output of the precomputation tree.
The size of the TL table is formulated as  
\begin{align}
\text{LUT}_{\text{TB}} = T \times Q \times N_{\text{TB}} \times \text{LUT}_{\text{entry}},  
\end{align}
where \( \text{LUT}_{\text{entry}} \) is the number of LUTs needed to store one entry in TL table. The lookup logic LUT consumption is given by  
\begin{align}
\text{LUT}_{\text{LPL}} = T \times Q \times \text{LUT}_{\text{lp}},
\end{align} 
where \( \text{LUT}_{\text{lp}} \) represents the LUT cost of a single lookup, its associated conversion logic, and reduction logic.
In the highly parallel \( T \times Q \) lookups in Methods 2, the dominant factor is \( \text{LUT}_{\text{LPL}} \), since the $N_{\text{TB}} \times \text{LUT}_{\text{entry}}$ is efficiently implemented using the distributed RAM architecture of primitives such as cascaded RAM32X1~\cite{AMD_UG974_ultrascale}. These argument are further validated in the ablation study  Sec.~\ref{SEC:ABLATION_TLMM}.

\captionsetup{font=small}
\captionsetup[subfigure]{font=footnotesize,labelformat=parens,labelsep=space}

\subsection{Fused Element-wise Operations}
Element-wise operations are computations performed on input tensors where each entry or entry pair is consumed only once. Within the Bitnet architecture, several such operations are critical: (1) quant from FP16 to INT8 and subsequent dequant from INT32 to FP16; (2) the channel-wise maximums for ABSMAX quantization; (3) the RoPE operator; (4) element-wise additions for residual connections; (5) element-wise multiplications in SwiGLU; and (6) the activation functions.
Previous research on LLM accelerators has overlooked the performance impact of these operations, often deeming them negligible \cite{moitra2025meadow}. In practical deployments, however, these operators can create performance bottlenecks, constraining the model's overall latency and resource utilization. This issue is acute in quantized modules, which involve a mix of efficient high-throughput INT computations and demanding FP operations. 
To address the above challenges, we introduce a specialized fusion unit designed to overlap the latency of these element-wise computations with the core TLMM operations, thereby optimizing execution efficiency. Moreover, a vector-wise operation is implemented to make sure we match the throughput of TLMM under resource constraints of DSP and LUT when designing FP operators. The detailed streaming pattern and datapath are displayed in Fig. \ref{fig:TLMM_FUSE}. The stream FIFO channels are in grey color with the type of vector it is buffering.

\subsubsection{INT8 Quantization and FP16 Dequantization}
To seamlessly fuse quantization with linear operations, a stream FIFO channel connects the respective functional units. To maximize data throughput, all FP16 values are packed into 256-bit vectors, each holding 16 elements. This vectorization enables parallel computation and efficient 256-bit wide AXI access to DDR memory. The quantization process involves a data-width transformation, converting these 16-element FP16 vectors into $T \times G$ INT8 vectors for the TLMM input FIFO, which necessitates stream resizing logic. Conversely, dequantization performs vectorized multiplication operations on the 16-element INT32 vectors to convert them to FP16 precision.


\subsubsection{Elementwise Activation-fused Multiplication and Addition}
The element-wise multiplication with SILU, element-wise addition, and RoPE operations are all performed in floating-point precision following dequantization. Both these three operations and bypass logic are implemented in TLMM-FUSE as displayed in Fig. \ref{fig:TLMM_FUSE} with an arbitrary unit deciding which submodule the stream FIFO flow in. As shown in Fig. \ref{Fig:Bitnet}, the element-wise multiplication is an integral component of the SwiGLU. This function operates on the outputs of the gate and up projections: the gate projection's output is first processed by a SILU activation, and the result is then multiplied element-wise with the up projection's output. The element-wise addition is applied immediately before the RMSNorm layer to incorporate the residual connection.


\subsubsection{RoPE Operation} 
For the RoPE implementation, we assume the necessary sinusoidal values are pre-computed and stored in DDR, as generating them on-the-fly would be resource- and time-intensive. The core RoPE function operates on a vector \( \mathbf{x} \in \text{FP16}^{d_h} \) by applying a rotation to pairs of elements. The mathematical formulation of the two implementations differs only in how the vector indices are grouped for these rotations:
\begin{equation}
\begin{cases}
\tilde{x}_m^{(2t)} = x_m^{(t)} \cos(m\theta_t) - x_m^{(t + d_h/2)} \sin(m\theta_t) \\
\tilde{x}_m^{(2t+1)} = x_m^{(t + d_h/2)} \cos(m\theta_t) + x_m^{(t)} \sin(m\theta_t)

\end{cases}\label{eq:half}
\end{equation}

\begin{equation}
\begin{cases}
\tilde{x}_m^{(2t)} = x_m^{(2t)} \cos(m\theta_t) - x_m^{(2t+1)} \sin(m\theta_t) \\
\tilde{x}_m^{(2t+1)} = x_m^{(2t+1)} \cos(m\theta_t) + x_m^{(2t)} \sin(m\theta_t)
\end{cases}\label{eq:consec}
\end{equation}
where \( 0 \leq t < d_h/2 \), \(m\) is the position, and the frequency basis \( \theta_t = 10000^{-2t/d_h} \).
As illustrated in Fig. \ref{fig:Rope}, the canonical LLaMA architecture~\cite{HuggingFaceLLaMAGithub} employs the interleaved pairing from eq. \ref{eq:half} . Although this pattern aligns well with the highly parallel and flexible memory systems of GPUs, it is fundamentally mismatched with streaming hardware architectures. In our FPGA-based streaming design, data flows through a function-level pipeline via FIFOs. The non-contiguous access pattern required by the interleaved pairing would force the pipeline to stall while buffering and gathering all elements for a given head.
By contrast, the consecutive pairing from eq. \ref{eq:consec} applied to data within a packed-vector is more hardware-friendly. This approach simplifies data access, reduces FIFO dependencies, and enables a lower initiation interval. To preserve mathematical equivalence with the LLaMA-style implementation while leveraging the benefits of consecutive access, a lossless transformation is necessary for query and key weight matrices. This transformation, applied on a per-head basis, follows the index exchange formula below, which converts the weight organization from the interleaved to the consecutive pattern:
\begin{align}
\begin{cases}
\mathbf{w}_{N_h}^{(2t)} \Leftrightarrow \mathbf{w}_{N_h}^{(t)} \\
\mathbf{w}_{N_h}^{(2t+1)} \Leftrightarrow \mathbf{w}_{N_h}^{(d_h/2 + t)}
\end{cases}
\quad \text{for } 0 \leq t < \frac{d_h}{2}.
\end{align}


\subsection{Weight Buffer Management Unit (WBMU)} \label{SEC:WBMU}
\subsubsection{Analytical Parameter Selection for TLMM based on Optimal URAM Utilization} \label{SEC:PARAMETER_WBMU}

In the design of LLM accelerators, external memory access for weights is a critical performance bottleneck, often dominating the latency of the decoding phase. This challenge is acute on edge FPGAs, where on-chip memory resources are limited. Among these resources, URAM is ideal for buffering large weight tensors.
Insufficient on-chip buffer capacity for weights leads to significant performance degradation. When the required weights cannot be stored on-chip URAM, the system must repeatedly access off-chip memory via the AXI bus, which will be severely limited by the low DDR bandwidth of the edge FPGA. 
Therefore, this work proposes an analytical model for optimizing URAM utilization by selecting the relevant TLMM parameters. This model provides a systematic approach to designing the on-chip weight buffer and TLMM accelerators, ensuring optimal parameter selection for both modules within the memory constraints of edge devices.

As introduced by the \cite{AMD_UG573_ultrascale_memory}, the size of the URAM resource is set as 72 bitwidth $\times$ 4096 depth, which can equal up to 16 BRAM18K blocks. However, due to its fixed bitwidth of 72, the design of storage architecture is vital. We assume there are $N_{\text{URAM}}$ URAM blocks in total in an edge FPGA. The total URAM memory size is $288Kb \times N_{\text{URAM}}$. In our design, each of the units saves a weight index vector $\bf{w}_{idx} \in \bf{W}_{idx}$ and the bit width is $T \times B_{\text{idx}}$. On one hand, our goal is to have a TL table as large as possible within the resource limitations; on the other hand we want to make sure the bitwidth of URAM is fully utilized. Thus, the largest number of TL table $T$ is set according to 
\begin{align}
T = \left\lceil \frac{72 \times c_{\text{URAM}}}{B_{\text{idx}}} \right\rceil
\label{eq:table},
\end{align}
where the factor $c_{\text{URAM}}$ stands for the cascade factor that combine $c_{\text{URAM}}$ URAM together to get a larger bitwidth. The $T$ is also confined by the available LUT resource to ensure place and route to implement the TLMM. The number of \text{LUT} can be utilized for TLMM is $\text{LUT}_{\text{max}}$. According to Sec. \ref{sec:TLMM_METHODS}, the constraint can be given by
\begin{equation}
\begin{split}
   T \times (N_{\text{TB}} \times \text{LUT}_{\text{tree}} + Q &\times N_{\text{TB}} \times 
   \text{LUT}_{\text{entry}} \\&
    + Q \times \text{LUT}_{\text{lp}} )  \leq \text{LUT}_{\text{max}},
\end{split}
\end{equation}
Moreover, to enable the parallel read of $Q$  trait, we will try to partition the target buffer array size of $X_{\text{URAM}} \times Y_{\text{URAM}}$ to ensure the above memory banks are independent.
Since each URAM is a dual-port memory ~\cite{AMD_UG573_ultrascale_memory}, we partition the weight matrix cyclically with a factor of $Q/2$. This strategy ensures that $Q$ consecutive elements are placed in independent blocks, enabling parallel access. 
These logical blocks are then mapped to physical URAMs based on the memory depth. The total number of utilized URAMs $U$ can be given as:
\begin{align}
U = \left\lceil \frac{(\frac {Y_{\text{URAM}}}{  Q} )\times \lceil \frac{X_{\text{URAM}}}{G \times T}\rceil}{4096}  \right\rceil  \times c_{\text{URAM}}  \times \frac{Q}{2}, \quad U \leq N_{\text{URAM}},
\end{align}
where the term wrapped by the ceil gives the depth required in total for a single partitioned array, $c_{\text{URAM}}$ is the cascade factor to build the required bitwidth, and there are $Q/2$ blocks in total. 

\subsubsection{TLMM Weight Accessing}
\begin{figure}[t]
    \centering
     \includegraphics[width=\linewidth]{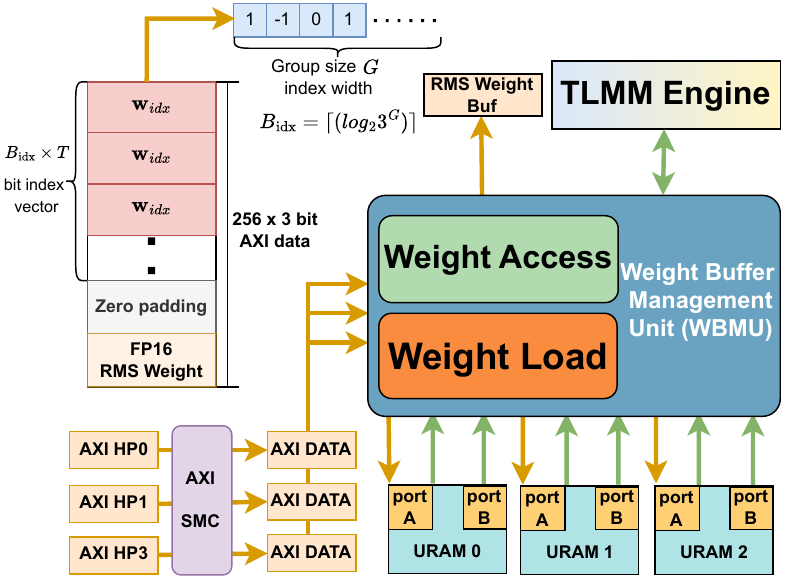}
     \vspace{-2mm}
    \caption{The WBMU and AXI data packing strategy. The green arrows denote the connection to the weight access, and the yellow arrows are connected to the weight load module.}
    \label{fig:WBMU_SYSTEM}
\end{figure}

A key challenge is supporting the various TLMM dimensions presented in Sec. \ref{sec:TLMM design}. The first two matmul types share the dimension $n = d_{\text{model}}$, so adapting to them primarily involves controlling memory access along the $k$-dimension. 
However, a complication arises because the weight matrices for the up- and down-projections are transposes of each other, and the address stride is supposed to be different. Furthermore, the dimension $d_{\text{ffn}}$ is not necessarily an even multiple of $d_{\text{model}}$. This dimensional mismatch prevents aligned memory access, which would necessitate complex logic to handle unaligned weight indices access. To resolve this, we pad the weight buffers to make their dimensions evenly divisible, thereby ensuring all memory accesses are aligned.
Moreover, the logical dimensions $d_{\text{ffn}}$ and $d_{\text{model}}$ are padded to be multiples of $T \times G$, resulting in the dimensions $d_{\text{ffn}}'$ and $d_{\text{model}}'$. Therefore, $X_{\text{URAM}} = d'_{\text{FFN}}$ and $Y_{\text{URAM}} = d'_{\text{model}}$.
The WBMU contains an address translation unit that converts a straightforward 2D software weight matrix index $(a, b)$ into a 3D-indexed physical address $(x, y, z)$. This scheme treats the cyclically partitioned URAM array as a single, contiguous address space for simplified indexing explanation. Its array size is configured as [$\left \lceil \frac{d_{\text{ffn}}'}{d_{\text{model}}'} \right \rceil$, $\frac{d_{\text{model}}'}{(T\times G)}$, $d_{\text{model}}$]. With weight index request $1 \leq a,b\leq d_{\text{ffn}}$ from TLMM engine, the address mapping is given by:
\begin{equation}
\begin{cases}
x = \begin{cases}
        0, &\text{q, k, v, o projection}\\
        \frac{b}{d_{\text{model}}} , &\text{Up projection} \\
        \frac{a}{  d_{\text{model}}'}, &\text{Down projection} \\
    \end{cases}\\

y = a \mod  d_{\text{model}}', \quad
z = b \mod d_{\text{model}}, \\
\end{cases}.
\end{equation}


\subsubsection{Off-chip DDR Weight Transfer}

As illustrated in Fig. \ref{fig:WBMU_SYSTEM}, the WBMU connects to the DDR memory system via three high performance (HP) ports: HP0, HP1, and HP3. The AXI interface for these ports is configured with a 256-bit data width, a burst size of 16, and an outstanding transaction limit of 16, following the guidelines in ~\cite{Benchmark_AXI_ON_ZYNQ}.
During weight loading, the three HP ports transmit data in parallel, as each is connected to an independent DDR Quality of Service (QoS) channel as shown in Fig. \ref{fig:sys_arch}. This parallelism results in a combined bandwidth of 768 bits per AXI access cycle. As shown in Fig. \ref{fig:WBMU_SYSTEM}(b), each 768-bit transfer is structured to carry $\lceil (768 - 16) / (T \times B)\rceil$ ${\bf w}_{\text{idx}}$ and one FP16 RMSNorm weight. In our design, with parameters $T=28$ and $B=5$, a single pipelined burst request can convey up to five weight index vectors and one RMSNorm weight in one transfer out of 16 burst AXI data. 

\subsection{RMS-MAX Unit}
The RMS-MAX hardware unit performs RMS normalization followed by channel-wise maximum finding for subsequent quantization and dequantization. Input data undergoes RMSnorm accumulation with upcasting to FP32 for precision, succeeded by FP16 division and FP16 RMSnorm weight scaling to compute the norm. A stream FIFO channel interfaces the RMSnorm output with the channel-wise max finding module: vector-wise maximums are first extracted per channel segment, and upon completion of the final channel flow, the global channel maximum is identified and stored in a dedicated buffer for quant/dequant operations. By decoupling the max-finding and quantization logic, memory dependencies are eliminated in computing the final quantized output. Furthermore, max-finding can proceed concurrently with vector-wise output FIFO due to their strictly sequential nature.
\captionsetup{font=small}
\captionsetup[subfigure]{font=footnotesize,labelformat=parens,labelsep=space}

\begin{figure*}[t]
\centering
\setlength{\abovecaptionskip}{4pt}
\setlength{\belowcaptionskip}{-4pt}
\setlength{\floatsep}{6pt}
\setlength{\textfloatsep}{6pt}
\begin{subfigure}{0.35\textwidth}
  \centering
  \includegraphics[width=\linewidth]{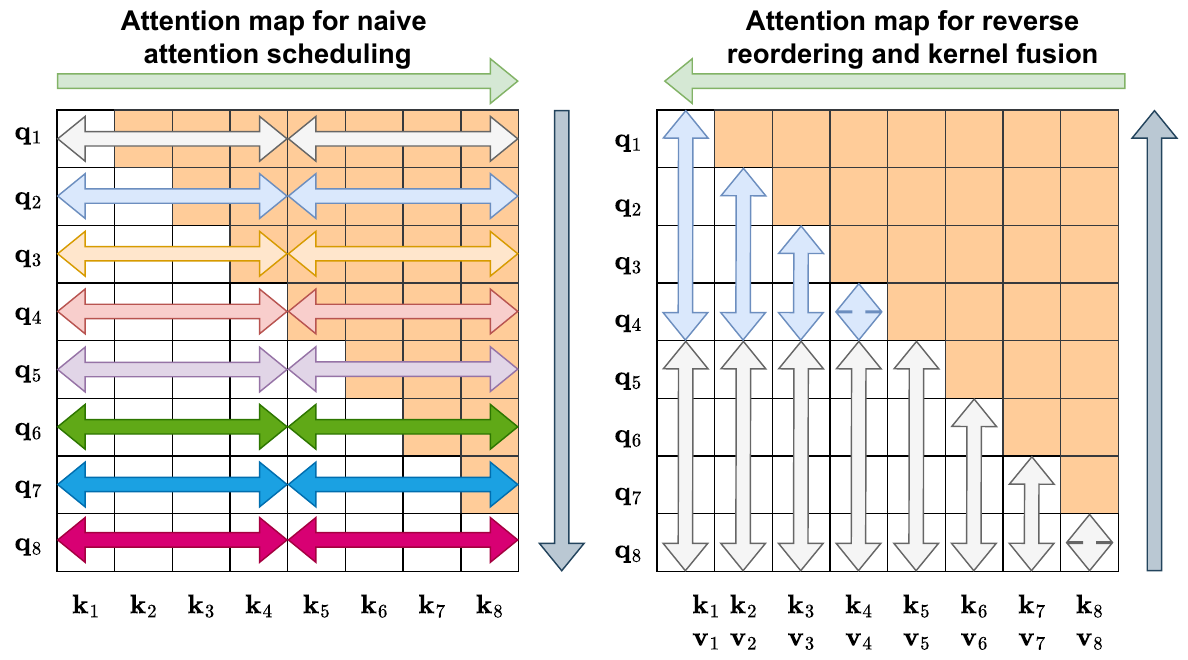}
  \caption{Schedule on the Attention Score Map ($N_{\text{pe}}{=}4$)}
  \label{fig:attention_map}
\end{subfigure}\hfill
\begin{subfigure}{0.29\textwidth}
  \centering
  \includegraphics[width=\linewidth]{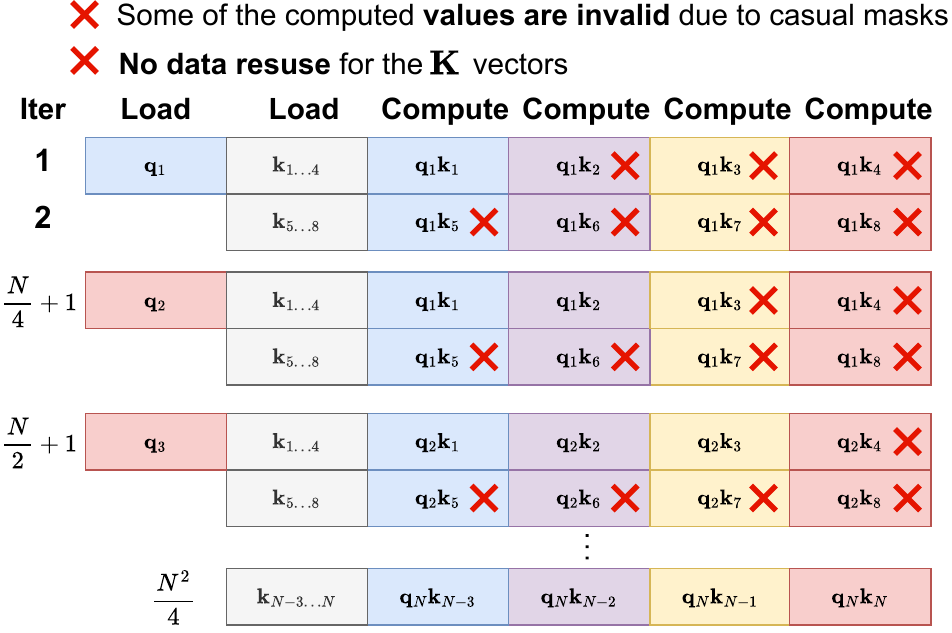}
  \caption{Naive Attention Scheduling ($N_{\text{pe}}{=}4$)}
  \label{fig:naive_schedule}
\end{subfigure}\hfill
\begin{subfigure}{0.35\textwidth}
  \centering
  \includegraphics[width=\linewidth]{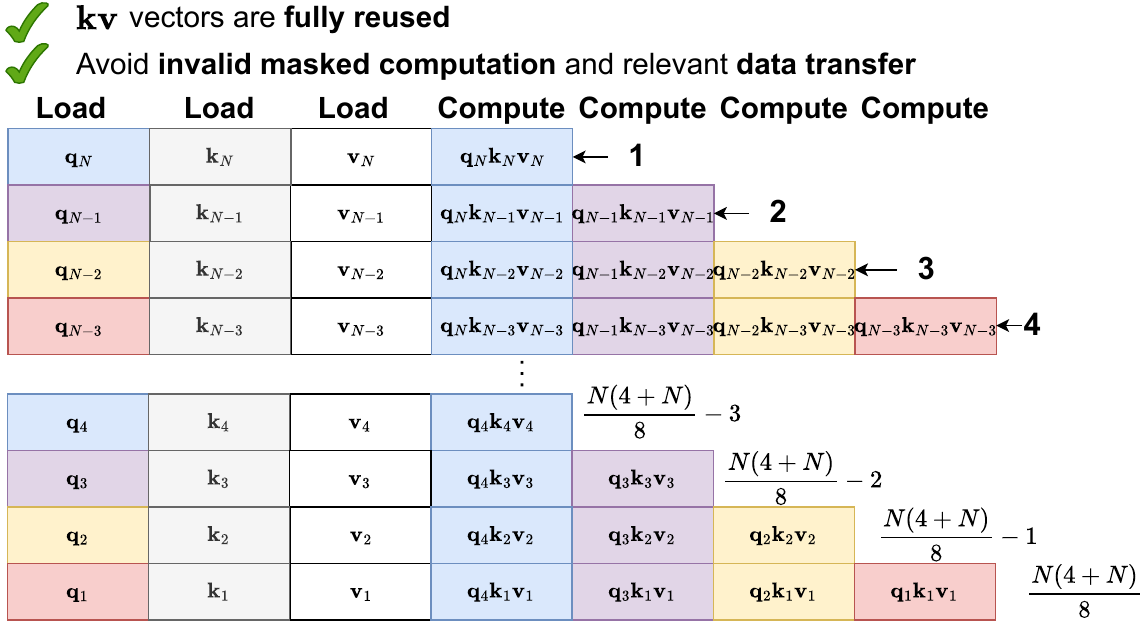}
  \caption{Reverse Attention Scheduling ($N_{\text{pe}}{=}4$)}
  \label{fig:reverse_schedule}
\end{subfigure}
 \caption{Attention Schedules Comparison}
\label{fig:attention_triptych}
\end{figure*}




\subsection{Reversed Prefill Attention Unit}
\subsubsection{Prefill Attention Challenge on edge FPGAs}
The prefill attention of LLM requires significant resources and bandwidth for multi-token computation at the sequence length of $N$, especially for attention computation, which involves softmax and matrix-to-matrix multi-head operations with a complexity of \( N^2 \). Given the limited memory bandwidth and finite computational units of DSP for FP operation on edge FPGAs, the computation order of prefill attention should be carefully scheduled to meet these requirements. Otherwise, it may be constrained by the bandwidth limitations of the edge FPGA, as shown in the naive attention scheduling in Fig. \ref{fig:naive_schedule} \cite{edgemoe}. When we have $N_{\text{pe}}$ processing elements (PEs), the bandwidth requirement will be $N_{\text{pe}}$ as well. The more PE we have, the larger the amount of data will be required from DDR, bounded by the total bandwidth. 


Furthermore, $N \times N \times N_{\text{head}}$ elements are stored back to DDR and loaded on chip as the softmax matrix $\bf S$.
The fusion of operations such as \( {\bf Q} {\bf K}^{\text{T}} \), softmax, and \( \bf S  {\bf V} \) can reduce the additional accesses to DDR. The state-of-the-art kernel fusion implementation for resource-abundant GPUs is Flash Attention \cite{dao2022flashattentionfastmemoryefficientexact}. However, GPU-optimized computation is not suitable for FPGAs, as GPUs have many more computational cores and much larger on-chip SRAM compared to the on-chip BRAM/URAM available on FPGAs. To address these challenges, we propose the reverse attention method, which utilizes fused attention and reverse reorder scheduling, specifically tailored for edge FPGAs.




\begin{figure}[t] 
\centering
\includegraphics[width=\linewidth]{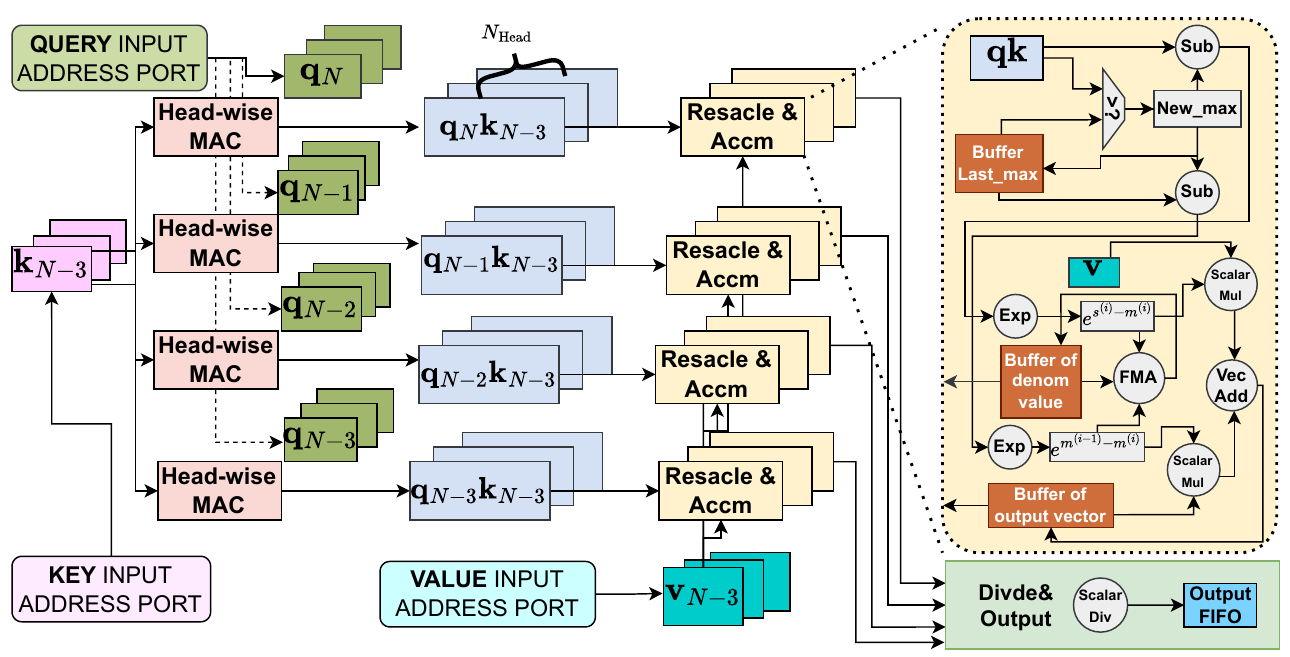}
\caption{Design of Fused Attention at Iteration Step 4 ($N_{\text{PE}} = 4$)}
\label{fig:reverse_pe} 
\vspace{1em}
\end{figure}

 \begin{figure}[ht]
    \centering

    \includegraphics[width=\linewidth]{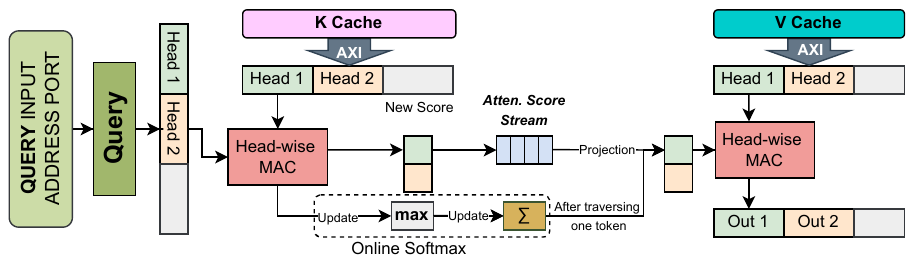}

    \caption{Detailed Design of Decoding Attention}
    \label{fig:decoding_atten}
\end{figure}
\subsubsection{Reversed and Fused Attention}


The reverse attention scheduling is depicted in Fig. \ref{fig:reverse_schedule}. Assume that the current sequence length of the prefill tokens is \( N \), with \( 1 < i \leq N \) and \( 1 < j \leq N \) representing the current token indices for \( {\bf q} \) and \( {\bf k}, {\bf v} \), respectively. There are a total of \( N_{{\text h}} \) heads. The operator of $\bf{qkv}$ denotes the fused operation.
The kernel fusion computation can be considered a special case of Flash Attention V2 ~\cite{dao2022flashattentionfastmemoryefficientexact} when the block size is equal to 1. The head-wise formula for the case with two consecutive blocks can be written as follows:
\begin{equation}
\vspace{-2mm}
\left\{
\begin{small}
\begin{aligned}
  s^{(i)} &= \frac{ {\bf q}_{i}{\bf k}^{\text{T}}_{i}}{\sqrt{d_{\text{h}}}}, m^{(i)} = \max\left(m^{(i-1)}, s^{(i)}\right)\\
\ell^{(i)} &= e^{m^{(i-1)} - m^{(i)}} \ell^{(i-1)} + e^{s^{(i)} - m^{(i)}} \\
\bf{o}^{(i)} &= e^{m^{(i-1)} - m^{(i)}}{ \bf o}^{(i-1)} + e^{s^{(i)} - m^{(i)}} {\bf v}_{i}
\end{aligned}
\end{small}
\right.
\end{equation}
where $m^{(i)}$ denotes the running maximum value, $s^{(i)}$ represents the dot product score, $\ell^{(i)}$ is the denominator factor, and ${\bf o}^{(i)}$ is the numerator vector. After all inputs have been processed (up to step N), the final attention output is computed as the division of the final numerator by the final denominator ${\bf o}^{(i)}/\ell^{(i)}$. 
Compared to a naive implementation, this online algorithm avoids generating the large intermediate attention score matrix $\bf S$, saving significant off-chip DDR access. It also breaks the data dependency on finding a global maximum value, which enables streaming computation and operation fusion.




Regarding detailed schedule, instead of starting from the first token \( {\bf q}_1 \), our schedule begins from \( {\bf q}_{N-1} \). Specifically, the level of parallelism is set to \( N_{\text{pe}} \). The factor \( N_{\text{pe}} \) also implies that the PE buffer can store \( N_{\text{pe}} \) tokens of \( {\bf q}_i \). In each iteration, one \( {\bf q}_i \) token is loaded onto the buffer of PEs with unicast (with the first batch loading \( {\bf q}_N \) to \( {\bf q}_{N-3} \)). Simultaneously, the corresponding \( {\bf k}_j \) and \( {\bf v}_j \) tokens are loaded to PEs in multicast. After all \( N \) \( {\bf k}_j \) and \( {\bf v}_j \) tokens have been loaded and the fused-kernel computation is completed, the next iteration will evict \( N_{\text{pe}} \) \( {\bf k}_j \) and \( {\bf v}_j \) tokens, starting from \( {\bf k}_{N-3} \) and \( {\bf v}_{N-3} \), to avoid redundant computations arising from the causal attention mask. Once all computation with regard to the loaded $\bf q$s is completed, the  ${\bf o}^{(i)}/\ell^{(i)}, i = N \text{ to } N-3$ for all heads are pushed to the as consecutive heads consist the hidden dimension. The microarchitecture state at step 4 is depicted in Fig.~\ref{fig:reverse_pe}. Each PE is divided into a head-wise multiplication and accumulation (MAC) unit, a rescale and accumulation unit for the output vector and denominator factor, and a final division and output unit.

The iteration continues until all \( 1 < i \leq N \), \( 1 < j \leq N \) are traversed. In this approach, the only required input buffers are for \( N_{\text{pe}} \) \( {\bf q}_i \) tokens, one \( {\bf k}_j \), and one \( {\bf v}_j \). Additionally, the intermediate buffers of the $N_{h}$ heads include: \( N_{h} \times N_{\text{pe}} \) multi-head MAC intermediate results \( s \), \( N_{h} \times N_{\text{pe}} \) multi-head previous max values \( m \), \( N_{h} \times N_{\text{pe}} \) intermediate denominators \( \ell \), as well as the output vector $\bf o$ buffer sized $d_{\text{model}} \times N_{\text{pe}}$. The trade-off also includes minor computational overhead: an additional exponentiation, $e^{m^{(i-1)} - m^{(i)}}$, is required to rescale previous results when the running maximum is updated. Additionally, the state of the running numerator $\bf{o}$ and denominator $\ell$ are maintained with two extra floating point rescales. 

To support the reverse reorder in the prefill stage, the embedding vectors are indexed by the reversed input prompt reordered token id, and the rope cache is also reversed. In this way, the RPA can be seamlessly utilized without extra reversed address DDR access logic that is not supported by the AXI address incremental burst access protocol \cite{amd_ug1037}. The advantage of the reversed attention is further proved in Sec. \ref{SEC:ABLATION_ATTENTION}.

\subsection{Decode Attention Unit}
The computational characteristics of attention differ between prefill and decoding phases. The computation in the decoding phase attention involves primarily matrix-vector and vector-vector operations. The computational load per step $O(Nd_{\text{model}})$ is significantly lower than the total matmul in prefill attention, $O(N^2d_{\text{model}})$. However, this phase requires fetching the large ${\bf K}_{\text{cache}}$ and ${\bf V}_{\text{cache}}$ matrices from memory (e.g., off-chip DDR) in every layer. Consequently, the decoding phase is often memory-bandwidth bound, especially as the sequence length $N$ grows. 
This heterogeneity of these workloads necessitates a system-level balance between performance and resource utilization to achieve optimal end-to-end efficiency.
For decoding attention, the marginal benefits of adding extra compute resources diminish compared with prefill attention. To address this, we decouple the fusion between $\bf{QK^{\text T}}$ multiplication (fetch ${\bf K}_{cache}$) and the weighted aggregation with ${\bf V}$ (fetch ${\bf V}_{cache}$), since these two operations are dominated by memory access. 
 Fig. \ref{fig:decoding_atten} illustrates the hardware architecture of our decoding attention design. The query input \textbf{Q} (single token) is pre-loaded on-chip, while the 
${\bf K}_{\text{cache}}$ and ${\bf V}_{\text{cache}}$ is loaded from off-chip DDR via AXI interface. We utilize a stream-like dataflow within the module to conceal the computation latency. We use an online softmax algorithm that partially fuses the softmax with $\bf{QK^{\text T}}$ vector-matrix multiplication. The intermediate attention score is stored in the on-chip stream buffer, so no off-chip memory access is required in softmax computation.

\begin{table*}[t]
\centering
\caption{Unified cross-platform and FPGA-based comparison for edge LLM inference. Resource utilization is reported for FPGA works. Throughput (TK/S) is tokens per second; energy efficiency (TK/J) is tokens per joule. Max DDR bandwidths are theoretical.}
\vspace{-2mm}
\label{tab:unified-llm-comparison}
\setlength{\tabcolsep}{3pt}
\renewcommand{\arraystretch}{0.85}
\resizebox{\textwidth}{!}{%
\begin{tabular}{c c c c c c c c c c c c c c c c}
\toprule
\multirow{2}{*}{\textbf{Work}} &
\multirow{2}{*}{\textbf{Platform}} &
\multirow{2}{*}{\textbf{Processor}} &
\multirow{2}{*}{\textbf{Model}} &
\multirow{2}{*}{\textbf{Precision}} &
\multirow{2}{*}{\textbf{Max DDR Bandwidth(GB/s)}} &
\multicolumn{5}{c}{\textbf{FPGA Resource Utilization}} &
\multirow{2}{*}{\textbf{Power (W)}} &
\multicolumn{2}{c}{\textbf{Throughput (TK/S)}} &
\multicolumn{2}{c}{\textbf{Energy Efficiency (TK/J)}} \\
\cmidrule(lr){7-11} \cmidrule(lr){13-14} \cmidrule(lr){15-16}
& & & & & & \textbf{LUT} & \textbf{FF} & \textbf{DSP} & \textbf{BRAM} & \textbf{URAM} & & \textbf{Prefill} & \textbf{Decode} & \textbf{Prefill} & \textbf{Decode} \\
\midrule
Raspberry Pi 5 \cite{adafruit2025qwen3} & SoC & $4\times$ \ Cortex-A76 & Qwen 0.6B & W4-A16 & 17.1 & — & — & — & — & — & 7.8 & 61.8 & 16.6 & 7.92 & 2.12 \\
Jetson Orin Nano \cite{jetsonailab2025slm} & GPU SoC & $8\times$ \ GPU SM & TinyLLaMA 1.1B & W4-A16 & 68.3 & — & — & — & — & — & 25 & 324.9 & 67.6 & 12.9 & 2.70 \\
\midrule
SECDA \cite{haris2024designing} & FPGA SoC & $2\times$ \ Cortex-A53 & TinyLLaMA 1.1B & W4-A16 & 2.1 & — & — & — & — & — & — & — & 0.6 & — & — \\
LLaMAF \cite{LLaMAf} & FPGA SoC & ZCU102 & TinyLLaMA 1.1B & W8-A8 & 19.2 & 164K & 171K & 528 & 223 & — & 5.1 & — & 1.5 & — & 0.29 \\
MEADOW \cite{moitra2025meadow} & FPGA SoC& ZCU102 & OPT 1.3B & W8-A8 & 19.2 & 150K & — & 845 & 2034 & — & 10 & 100 & 2 & 10 & 0.20 \\
\midrule
\textbf{TeLLMe (KV260)} & FPGA SoC & — & BitNet 0.73B & W1.58-A8 & 17.1 & 98K & 137K & 610 & 98.5 & 60 & \textbf{4.8} & \textbf{143} & \textbf{25} & \textbf{29.8} & \textbf{5.2} \\
\bottomrule
\end{tabular}
}
\end{table*}

\begin{table}[t]
\centering
\caption{Perplexity and Intelligence/J comparison on WikiText-2. Intelligence/J is defined as $\frac{tokens/s}{(PPL \cdot Power)}$~\cite{tenent}. Throughput (TK/S) is tokens per second.}
\vspace{-2mm}
\label{tab:ppl-intelligenceJ}
\setlength{\tabcolsep}{4pt}
\renewcommand{\arraystretch}{0.85}
\resizebox{0.48\textwidth}{!}{%
\begin{tabular}{c c c c c c c c}
\toprule
\multirow{2}{*}{\textbf{Work \& Platform}} &
\multirow{2}{*}{\textbf{Model}} &
\multirow{2}{*}{\textbf{Power (W)}} &
\multirow{2}{*}{\textbf{WT-2 (PPL)}} &
\multicolumn{2}{c}{\textbf{Throughput (TK/S)}} &
\multicolumn{2}{c}{\textbf{Intelligence/J}} \\
\cmidrule(lr){5-6} \cmidrule(lr){7-8}
& & & & \textbf{Prefill} & \textbf{Decode} & \textbf{Prefill} & \textbf{Decode} \\
\midrule
Raspberry Pi 5 & Qwen 0.6B & 7.8  & 24.00  & 61.8   & 16.6  & 0.330 & 0.089 \\
Jetson Orin Nano & TinyLLaMA 1.1B & 25.0 & 12.42  & 324.9  & 67.6  & 1.046 & 0.218 \\
SECDA (Pynq) & TinyLLaMA 1.1B & 1.2    & 12.42  & —      & 0.6   & —     & 0.040  \\
LLaMAF (ZCU102) & TinyLLaMA 1.1B & 5.1    & 8.89 & —      & 1.5   & —     & 0.041     \\
MEADOW (ZCU102) & OPT 1.3B & 10.0 & 15.41  & 100.0  & 2.0   & 0.649 & 0.013 \\
\textbf{TeLLMe (KV260)} & \textbf{BitNet 0.73B} & \textbf{4.8} & \textbf{12.79} & \textbf{143} & \textbf{25} & \textbf{2.330} & \textbf{0.407} \\
\bottomrule
\end{tabular}
 }
\end{table}
\section{Experimental Results and Analysis} \label{expr}
\subsection{Experiment Setup}
We implement the TeLLMe accelerator using high-level synthesis in C/C++ with Vitis HLS 2024.1 and Vivado 2024.1. The design is evaluated on the AMD Kria KV260 platform (Zynq UltraScale+ XCK26 MPSoC). To achieve timing closure, a clock frequency of 250 MHz is employed for the final bitstream generation. The end-to-end latency results are measured with the PYNQ Runtime library, while the latency breakdown is derived from cycle-accurate RTL simulations. For fair comparison, we select prior works that implement approximately 1B-parameter models on edge platforms.
The implemented model is BitNet 0.73B~\cite{bitnet158}, comprising 49M parameters for the language model head weights and embedding token table, alongside 680M parameters for the Transformer decoder layer weights.
According to the parameter selection formulas in Sec.~\ref{SEC:PARAMETER_WBMU}, the parameters for our harderware accelerator are $G = 3$, $T = 28$, $Q = 16$, $U = 48$, $B_{\text{idx}} = 5$, with the weight index vector ${\bf w}_{\text{idx}}$ having a 140-bit width.
Limited by routing and placement constraints for an extra module, we offload the language model head (LM Head) computation to the Processing System (PS) side. Using ARM NEON SIMD computing kernel with W8A8 quantization, the LM Head latency is 9 ms, with negligible accuracy loss, as shown in Table~\ref{tab:ppl-intelligenceJ}. This latency is considered in our end-to-end performance results.

\begin{figure}[h]
    \centering
    \includegraphics[width=0.95\linewidth]{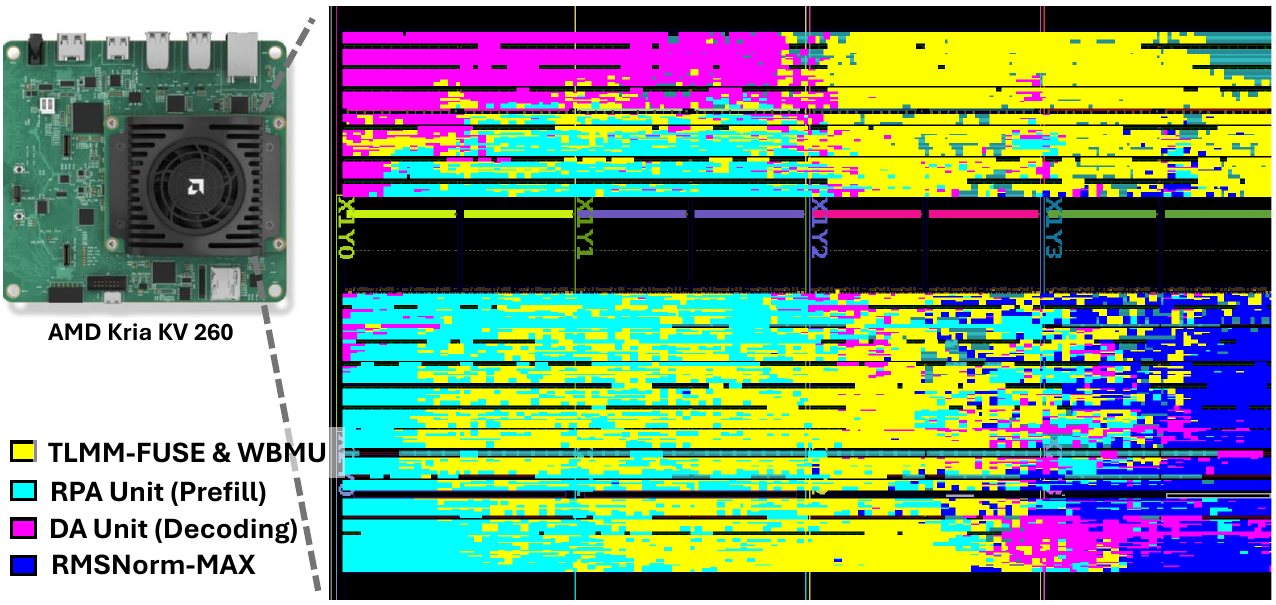}
    \caption{TeLLMe Hardware Accelerator Floorplan on KV260}
    \label{fig:device_map}
\end{figure}


\subsection{TeLLMe Inference Performance}
\begin{figure}[h]
    \centering

    \includegraphics[width=\linewidth]{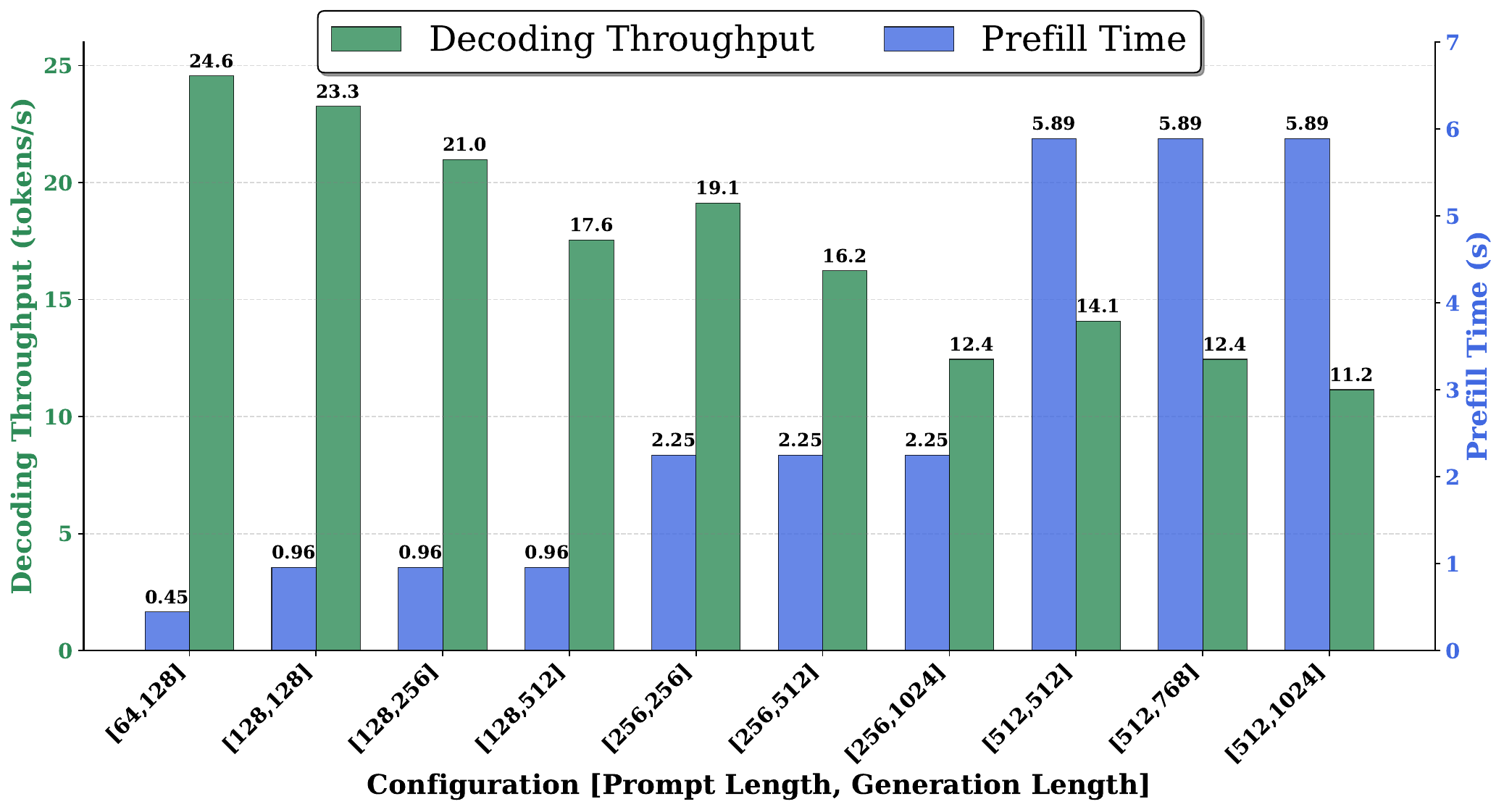}

    \caption{TeLLMe LLM Inference Performance}
    \label{fig:perf}
\end{figure}

Fig.~\ref{fig:perf} presents the performance characteristics of our TeLLMe accelerator across ten different [prompt, generation] configurations. The results demonstrate a clear inverse relationship between sequence length and performance metrics. Our design achieves up to 25 tokens/s throughput with great real word prefill latency. 

Notably, configurations with prompt lengths less than 256 tokens achieve decoding throughput above 16 tokens/s and prefill latency below 2.25s, demonstrating TeLLMe's performance under limited resource \& power budget. These results also suggest that practical FPGA deployments should target moderate sequence lengths to balance latency requirements with inference throughput, achieving optimal efficiency ratios exceeding 20 tokens/s per prefill second for configurations [64,128] through [128,256].

\begin{figure}[ht]
    \centering

    \includegraphics[width=\linewidth]{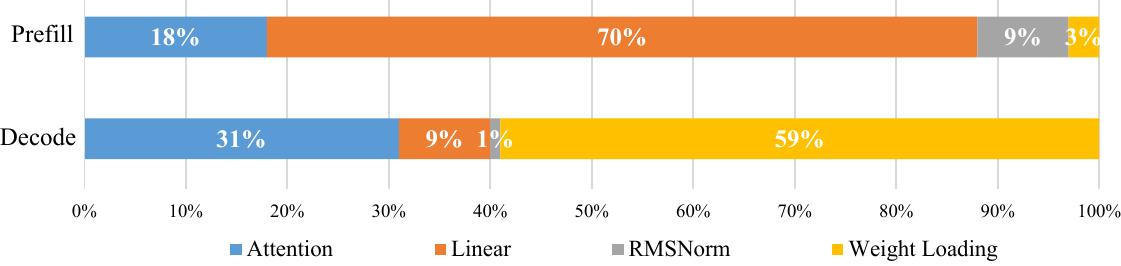}

    \caption{TeLLMe LLM Inference Latency Breakdown (Length = 128)}
    \label{fig:breakdown}
\end{figure}

Our inference latency breakdown in Fig.~\ref{fig:breakdown} reveals distinct computational characteristics between the prefill and decoding phases on FPGA. The decoding phase is predominantly memory-bound, with performance limited by memory bandwidth due to weight loading and KV cache accesses, while the prefill phase is compute-bound, requiring substantial parallel matrix operations. By leveraging the KV cache, decoding attention processes only a single new token against cached keys and values, consuming significantly less time per token than prefill, which recomputes attention across the entire sequence. This fundamental difference motivates our architecture design.
The further resource breakdown is shown in Table \ref{tab:utilization}. 


\begin{table}[h]
\centering
\caption{FPGA resource consumption breakdown. Percentages at bottom indicate design utilization per resource.}
\label{tab:utilization}
\vspace{-2mm}
\setlength{\tabcolsep}{5pt}
\renewcommand{\arraystretch}{0.8}
\resizebox{\columnwidth}{!}{%
\begin{tabular}{lrrrrr}
\toprule
\textbf{Module} & \textbf{LUT} & \textbf{FF} & \textbf{BRAM} & \textbf{URAM} & \textbf{DSP} \\
\midrule
TLMM\texttt{-}FUSE Unit          & 43{,}137 & 51{,}894 &  5.5 &  0  & 320 \\
RPA Unit (Prefill Attention)        & 18{,}072 & 31{,}212 & 43.5 &  4  & 115 \\
DA Unit (Decoding Attention)       &  9{,}570 & 20{,}138 & 11.5 &  4  & 123 \\
RMS-MAX Unit                          &  6{,}092 & 11{,}175 &  4.0 &  4  &  47 \\
WBMU Unit                   &  6{,}885   &  1{,}614  & —    &  48   &  — \\
Control \& Communication                  & 9{,}114 & 13{,}412 & 34.0 & —  &   5 \\
Misc                             &  5{,}433 &  7{,}276 &   —  &  —  &   — \\
\midrule
\textbf{Total}                   & \textbf{98{,}303} & \textbf{136{,}721} & \textbf{98.5} & \textbf{60} & \textbf{610} \\
\textit{Utilization}             & \textit{(84\%)}   & \textit{(28\%)}     & \textit{(68\%)} & \textit{(94\%)} & \textit{(49\%)} \\
\bottomrule
\end{tabular}%
}
\vspace{-2mm}
\end{table}
\begin{table}[h]
\caption{LUT usage breakdown for different TLMM Designs}
\vspace{-2mm}
\label{tab:LUT}
\vspace{-2mm}
\centering
\setlength{\tabcolsep}{5pt}
\renewcommand{\arraystretch}{0.95}
\resizebox{\columnwidth}{!}{%
\begin{tabular}{lrrrr}
\toprule
\textbf{Design} & $\textbf{LUT}_{\text{PRE}}$ & $\textbf{LUT}_{\text{TL}}$ & $\textbf{LUT}_{\text{LPL}}$ & \textbf{TOTAL} \\
\midrule
None Table Lookup (Method 1) & -- & -- &  -- & 43,176 \\
Partial Storage Table Lookup (Method 2)& 3,117 & 6,440 & 25,643 & 35,200 \\
Full Storage Table Lookup (Ours) & 5,301  & 11,452 & 6,329 & 23,082 \\
\bottomrule
\end{tabular}
}
\end{table}

\subsection{Cross-Platform and FPGA-based Comparison for Edge LLM Inference}
We present a comprehensive evaluation of TeLLMe against state-of-the-art edge LLM accelerators across FPGA, CPU, and GPU platforms. Table~\ref{tab:unified-llm-comparison} summarizes the hardware characteristics and performance metrics, while Table~\ref{tab:ppl-intelligenceJ} provides quality-aware efficiency comparisons using the Intelligence/J metric, defined as throughput normalized by perplexity and power consumption.

\subsubsection{Comparison with FPGA-based Solutions}

TeLLMe demonstrates substantial advantages over existing FPGA implementations in both throughput and energy efficiency. Compared to SECDA, which achieves 0.6 tokens/s decoding throughput on a Pynq-Z2 board, TeLLMe delivers 41.7$\times$ higher decoding throughput (25 tokens/s) while maintaining comparable power consumption (4.8W). This significant performance gain stems from our optimized BitNet architecture and efficient memory subsystem design that better exploits the available memory bandwidth.

When compared to LLaMAF on the higher-end ZCU102 platform, TeLLMe achieves 16.7$\times$ higher decoding throughput despite running on the more resource-constrained KV260 SoC. Although LLaMAF benefits from the ZCU102's superior memory bandwidth (19.2 GB/s vs 17.1 GB/s) and operates at slightly higher power (5.1W), TeLLMe's energy efficiency is 17.9$\times$ better during decoding operations. This improvement is primarily attributed to BitNet's 1.58-bit weight quantization, which dramatically reduces memory traffic compared to conventional 8-bit quantization.

MEADOW implements a well-designed INT8 matmul and attention engine. However, it lacks detailed descriptions of FP operations, including quantization and dequantization processes, to ensure smooth quantization support. Meanwhile, it represents a more computation-intensive approach, deploying a larger OPT 1.3B model with extensive DSP utilization (845 DSPs) despite delivering worse perform model (15.41 vs TeLLMe 12.79 of WikiText-2 PPL). While MEADOW achieves fair prefill throughput (100 tokens/s), its decoding performance (2 tokens/s) is 12.5$\times$ lower than TeLLMe. More critically, TeLLMe achieves 26$\times$ higher decoding energy efficiency (5.2 TK/J vs 0.20 TK/J), demonstrating the effectiveness of our memory-optimized architecture for the memory-bound decoding phase. The Intelligence/J metric further highlights this advantage: TeLLMe achieves 31.3$\times$ better decoding intelligence efficiency (0.407 vs 0.013), indicating superior quality-normalized energy efficiency. 

From a resource utilization perspective, TeLLMe achieves these performance gains while maintaining modest FPGA resource consumption. Our design uses 98K LUTs and 610 DSPs, comparable to or lower than existing solutions, while leveraging URAMs (60 blocks) for efficient on-chip weight storage. This balanced resource allocation enables deployment on mid-range FPGA SoCs without requiring high-end devices. 

It is worth noting that both LLaMAF and MEADOW focus exclusively on accelerating the decoding stage, assuming that prefill operations will be offloaded to separate devices such as host CPUs. In contrast, TeLLMe provides a unified acceleration framework that efficiently handles both prefill and decoding phases on a single FPGA device. This holistic approach eliminates the need for heterogeneous processing pipelines and associated data transfer overhead. Despite accelerating both inference stages, TeLLMe uses fewer hardware resources while delivering superior throughput in both phases, achieving 143 tokens/s prefill and 25 tokens/s decoding performance. This comprehensive acceleration capability, combined with leading energy efficiency across all metrics, establishes TeLLMe as the state-of-the-art solution for FPGA-based edge LLM inference.

\subsubsection{Comparison with CPU and GPU Edge Platforms}
While GPU and CPU platforms offer higher absolute throughput due to greater computational resources and memory bandwidth, TeLLMe provides compelling advantages in energy efficiency and cost-effectiveness. The Jetson Orin Nano, equipped with 8 GPU streaming multiprocessors (SM) and 68.3 GB/s memory bandwidth, achieves 67.6 tokens/s decoding throughput at 25W power consumption. In contrast, TeLLMe delivers 25 tokens/s at only 4.8W, resulting 1.9$\times$ better decoding energy efficiency despite the GPU's 4$\times$ higher memory bandwidth, highlighting the benefits of BitNet's reduced memory access requirements.

The Raspberry Pi 5, representing a CPU-based edge solution, operates at 7.8W and achieves 16.6 tokens/s decoding throughput. TeLLMe surpasses this performance with 1.5$\times$ higher throughput while consuming 38\% less power, resulting in 2.45$\times$ better decoding energy efficiency. This comparison is particularly relevant for cost-sensitive edge deployments where both bill-of-materials cost and operational power budgets are critical constraints.

When considering the Intelligence/J metric, TeLLMe achieves the highest decoding intelligence efficiency (0.407) among all evaluated platforms, representing 1.9$\times$ improvement over the Jetson Orin Nano (0.218) and 4.6$\times$ improvement over the Raspberry Pi 5 (0.089). This metric accounts for both model quality (perplexity) and energy consumption, providing a holistic measure of inference efficiency. The superior Intelligence/J ratio validates that TeLLMe's combination of BitNet's compact representation and FPGA-optimized architecture delivers excellent inference quality per unit of energy, making it well-suited for edge applications where battery life and thermal constraints are paramount.

Furthermore, TeLLMe's perplexity (12.79) is competitive with larger models like TinyLLaMA-1.1B-W4A16 (12.42) while operating at significantly lower power. This demonstrates that BitNet's aggressive quantization maintains model quality effectively, enabling efficient edge deployment without substantial accuracy degradation. The comparison with LLaMAF's perplexity (8.89) shows room for model quality improvement, which we identify as future work through additional training sample and BitNet model architectural refinements.

\subsection{Ablation Study}
\subsubsection{Effect of TLMM Design Choice}\label{SEC:ABLATION_TLMM}
As for comparison, the LUT consumption of different TLMM unit design methods presented in \ref{sec:TLMM_METHODS} is given. The configuration for the TLMM for ablation study is set as \( G = 3 \), \( T = 28 \), and \( Q = 16 \) to guarantee the same level of parallelism and latency. Our design of the TLMM core consumes 23,082 LUTs, while the naive selection (Method 1), which selects whether to add or subtract, requires 43,176 LUTs. Another approach (Method 2) involves storing half of the possible combinations (13 out of 27) instead of all combinations and using the index to determine whether the value should be negative. This approach results in a smaller distributed RAM size, aiming to save LUTs. However, after synthesizing, it consumes 35,200 LUTs in parallel logic in $T\times G$ looking up that inverts the bit when the value is identified to be negative. The detail decompostion of the LUT usage is displayed in Table \ref{tab:LUT}, which supports the fact that $\text{LUT}_{\text{LPL}}$ is dominant in LUT consumption.

\subsubsection{Effect of Prefill Attention Design Choice}\label{SEC:ABLATION_ATTENTION}
To demonstrate the efficacy of the reverse attention unit, we implemented a baseline naive attention design as depicted in Fig.~\ref{fig:naive_schedule}. For a fair comparison, both designs employed an identical number of processing elements, with $N_{\text{pe}} = 8$. Our reverse attention design utilized 115 DSPs, compared to 105 for the naive implementation, owing to the additional exponential operations. However, in on-board latency measurements at a prefill sequence length of 128, our design achieved a latency of 7.6 ms, whereas the naive attention required 14.3 ms. This performance improvement validates the advantages of our module on bandwidth-constrained edge FPGAs. 

\section{Conclusion}
In this work, we presented TeLLMe, the first end-to-end FPGA accelerator for ternary LLMs. TeLLMe accelerates both prefill and decoding stages. 
We propose to integrate the table-lookup-based matrix multiplication into the TeLLMe accelerator, achieving up to 25 tokens/s generation throughput and up to 143 tokens/s prefill throughput while consuming under 5W power budget and maintaining great model performance. TeLLMe achieves a significant performance and efficiency improvement over existing mobile-edge devices and previous FPGA-based accelerators, 
enabling energy-efficient, low-latency generative AI in the edge environment.


\bibliographystyle{unsrt}
\bibliography{references}

\begin{thebibliography}{10}

\bibitem{brown2020language}
Tom Brown, Benjamin Mann, Nick Ryder, Melanie Subbiah, Jared~D Kaplan, Prafulla Dhariwal, Arvind Neelakantan, Pranav Shyam, Girish Sastry, Amanda Askell, et~al.
\newblock Language models are few-shot learners.
\newblock {\em Advances in neural information processing systems}, 33:1877--1901, 2020.

\bibitem{touvron2023LLaMA}
Hugo Touvron, Thibaut Lavril, Gautier Izacard, Xavier Martinet, Marie-Anne Lachaux, Timoth{\'e}e Lacroix, Baptiste Rozi{\`e}re, Naman Goyal, Eric Hambro, Faisal Azhar, et~al.
\newblock Llama: Open and efficient foundation language models.
\newblock {\em arXiv preprint arXiv:2302.13971}, 2023.

\bibitem{guo2025deepseek}
Daya Guo, Dejian Yang, Haowei Zhang, Junxiao Song, Ruoyu Zhang, Runxin Xu, Qihao Zhu, Shirong Ma, Peiyi Wang, Xiao Bi, et~al.
\newblock Deepseek-r1: Incentivizing reasoning capability in llms via reinforcement learning.
\newblock {\em arXiv preprint arXiv:2501.12948}, 2025.

\bibitem{qiao2022two}
Ye~Qiao, Mohammed Alnemari, and Nader Bagherzadeh.
\newblock A two-stage efficient 3-d cnn framework for eeg based emotion recognition.
\newblock In {\em 2022 IEEE International Conference on Industrial Technology (ICIT)}, pages 1--8. IEEE, 2022.

\bibitem{10025006}
Andrew Ding, Ye~Qiao, and Nader Bagherzadeh.
\newblock Bnn an ideal architecture for acceleration with resistive in memory computation.
\newblock {\em IEEE Transactions on Emerging Topics in Computing}, 11(2):281--291, 2023.

\bibitem{bitnet}
Hongyu Wang, Shuming Ma, Li~Dong, and et~al.
\newblock Bitnet: Scaling 1-bit transformers for large language models.
\newblock {\em arXiv preprint arXiv:2310.11453}, 2023.

\bibitem{qiao2025cobra}
Ye~Qiao, Zhiheng Chen, Yian Wang, Yifan Zhang, Yunzhe Deng, and Sitao Huang.
\newblock Cobra: Algorithm-architecture co-optimized binary transformer accelerator for edge inference.
\newblock {\em arXiv preprint arXiv:2504.16269}, 2025.

\bibitem{qiao2025tellme}
Ye~Qiao, Zhiheng Chen, Yifan Zhang, Yian Wang, and Sitao Huang.
\newblock Tellme: An energy-efficient ternary llm accelerator for prefilling and decoding on edge fpgas.
\newblock {\em arXiv preprint arXiv:2504.16266}, 2025.

\bibitem{bitnet158}
Shuming Ma, Hongyu Wang, Lingxiao Ma, and et~al.
\newblock The era of 1-bit llms: All large language models are in 1.58 bits.
\newblock {\em arXiv preprint arXiv:2402.17764}, 2024.

\bibitem{deepseek}
Daya Guo, Dejian Yang, Haowei Zhang, and et~al.
\newblock Deepseek-r1: Incentivizing reasoning capability in llms via reinforcement learning.
\newblock {\em arXiv preprint arXiv:2501.12948}, 2025.

\bibitem{li2025pushing}
Jindong Li, Tenglong Li, Guobin Shen, Dongcheng Zhao, Qian Zhang, and Yi~Zeng.
\newblock Pushing up to the limit of memory bandwidth and capacity utilization for efficient llm decoding on embedded fpga.
\newblock {\em arXiv preprint arXiv:2502.10659}, 2025.

\bibitem{LUTNET}
Erwei Wang, James~J. Davis, Peter Y.~K. Cheung, and George~A. Constantinides.
\newblock Lutnet: Rethinking inference in fpga soft logic.
\newblock In {\em 2019 IEEE 27th Annual International Symposium on Field-Programmable Custom Computing Machines (FCCM)}, pages 26--34, 2019.

\bibitem{SUMLUTNET}
Olivia Weng, Marta Andronic, Danial Zuberi, Jiaqing Chen, Caleb Geniesse, George~A. Constantinides, Nhan Tran, Nicholas~J. Fraser, Javier~Mauricio Duarte, and Ryan Kastner.
\newblock Greater than the sum of its luts: Scaling up lut-based neural networks with amigolut.
\newblock In {\em Proceedings of the 2025 ACM/SIGDA International Symposium on Field Programmable Gate Arrays}, FPGA '25, page 25–35, New York, NY, USA, 2025. Association for Computing Machinery.

\bibitem{TLMAC}
Daniel Gerlinghoff, Benjamin Choong, Rick Goh, Weng-Fai Wong, and Tao Luo.
\newblock Table-lookup mac: Scalable processing of quantised neural networks in fpga soft logic.
\newblock pages 235--245, 04 2024.

\bibitem{tmac}
Jianyu Wei, Shijie Cao, Ting Cao, and et~al.
\newblock T-mac: Cpu renaissance via table lookup for low-bit llm deployment on edge.
\newblock {\em arXiv preprint arXiv:2407.00088}, 2024.

\bibitem{LLaMAf}
Han Xu, Yutong Li, and Shihao Ji.
\newblock Llamaf: An efficient llama2 architecture accelerator on embedded fpgas.
\newblock {\em arXiv preprint arXiv:2409.11424}, 2024.

\bibitem{edgemoe}
Rishov Sarkar, Hanxue Liang, Zhiwen Fan, Zhangyang Wang, and Cong Hao.
\newblock Edge-moe: Memory-efficient multi-task vision transformer architecture with task-level sparsity via mixture-of-experts.
\newblock In {\em IEEE/ACM International Conference on Computer-Aided Design (ICCAD)}, pages 1--9, 2023.

\bibitem{haris2024designing}
Jude Haris, Rappy Saha, Wenhao Hu, and José Cano.
\newblock Designing efficient llm accelerators for edge devices.
\newblock {\em arXiv preprint arXiv:2408.00462}, 2024.
\newblock Accessed: 2025-04-20.

\bibitem{moitra2025meadow}
Abhishek Moitra, Arkapravo Ghosh, Shrey Agarwal, Aporva Amarnath, Karthik Swaminathan, and Priyadarshini Panda.
\newblock Meadow: Memory-efficient dataflow and data packing for low power edge llms.
\newblock {\em arXiv preprint arXiv:2503.11663}, feb 2025.

\bibitem{xiao2023smoothquant}
Guangxuan Xiao, Ji~Lin, Mickael Seznec, Hao Wu, Julien Demouth, and Song Han.
\newblock {S}mooth{Q}uant: Accurate and efficient post-training quantization for large language models.
\newblock In {\em Proceedings of the 40th International Conference on Machine Learning}, 2023.

\bibitem{LightMamba}
Renjie Wei, Songqiang Xu, Linfeng Zhong, Zebin Yang, Qingyu Guo, Yuan Wang, Runsheng Wang, and Meng Li.
\newblock Lightmamba: Efficient mamba acceleration on fpga with quantization and hardware co-design.
\newblock In {\em 2025 Design, Automation \& Test in Europe Conference (DATE)}, pages 1--7, 2025.

\bibitem{FlightLLM}
Shulin Zeng, Jun Liu, Guohao Dai, Xinhao Yang, Tianyu Fu, Hongyi Wang, Wenheng Ma, Hanbo Sun, Shiyao Li, Zixiao Huang, Yadong Dai, Jintao Li, Zehao Wang, Ruoyu Zhang, Kairui Wen, Xuefei Ning, and Yu~Wang.
\newblock Flightllm: Efficient large language model inference with a complete mapping flow on fpgas.
\newblock In {\em Proceedings of the 2024 ACM/SIGDA International Symposium on Field Programmable Gate Arrays}, FPGA '24, page 223–234, New York, NY, USA, 2024. Association for Computing Machinery.

\bibitem{EdgeLLM}
Mingqiang Huang, Ao~Shen, Kai Li, Haoxiang Peng, Boyu Li, Yupeng Su, and Hao Yu.
\newblock Edgellm: A highly efficient cpu-fpga heterogeneous edge accelerator for large language models.
\newblock {\em IEEE Transactions on Circuits and Systems I: Regular Papers}, 2025.

\bibitem{yin2025tereffic}
Chenyang Yin, Zhenyu Bai, Pranav Venkatram, Shivam Aggarval, Zhaoying Li, and Tulika Mitra.
\newblock Tereffic: Highly efficient ternary llm inference on fpga.
\newblock {\em arXiv preprint arXiv:2502.16473}, 2025.

\bibitem{AMDVersalOverview}
{AMD}.
\newblock Versal adaptive socs - {AMD}.
\newblock \url{https://www.amd.com/en/products/adaptive-socs-and-fpgas/versal.html}, 2024.
\newblock Accessed: 2024-01-20.

\bibitem{LOGICNET}
Yaman Umuroglu, Yash Akhauri, Nicholas~James Fraser, and Michaela Blott.
\newblock Logicnets: Co-designed neural networks and circuits for extreme-throughput applications.
\newblock In {\em 2020 30th International Conference on Field-Programmable Logic and Applications (FPL)}, pages 291--297, 2020.

\bibitem{AMD_UG974_ultrascale}
{Advanced Micro Devices}.
\newblock Ultrascale architecture libraries guide.
\newblock Technical Report UG974, Advanced Micro Devices, May 2024.
\newblock Version 2024.1. Describes UltraScale primitives such as RAM32X1D.

\bibitem{HuggingFaceLLaMAGithub}
Hugging Face.
\newblock transformers/models/llama at main · huggingface/transformers, 2023.
\newblock GitHub repository.

\bibitem{AMD_UG573_ultrascale_memory}
{Advanced Micro Devices}.
\newblock Ultrascale architecture memory resources user guide.
\newblock Technical Report UG573, Advanced Micro Devices, September 2021.
\newblock Version 1.13.

\bibitem{Benchmark_AXI_ON_ZYNQ}
Kristiyan Manev, Anuj Vaishnav, and Dirk Koch.
\newblock Unexpected diversity: Quantitative memory analysis for zynq ultrascale+ systems.
\newblock In {\em 2019 International Conference on Field-Programmable Technology (ICFPT)}, pages 179--187, 2019.

\bibitem{dao2022flashattentionfastmemoryefficientexact}
Tri Dao, Daniel~Y. Fu, Stefano Ermon, Atri Rudra, and Christopher Ré.
\newblock Flashattention: Fast and memory-efficient exact attention with io-awareness, 2022.

\bibitem{amd_ug1037}
{Advanced Micro Devices}.
\newblock {\em {Vivado Design Suite: AXI Reference Guide (UG1037)}}.
\newblock Advanced Micro Devices, 12 2022.
\newblock Version 2022.2.

\bibitem{adafruit2025qwen3}
{Adafruit}.
\newblock Local llms on raspberry pi: Qwen3.
\newblock \url{https://learn.adafruit.com/local-llms-on-raspberry-pi/qwen3}, 2025.
\newblock Accessed: 2025-9-01.

\bibitem{jetsonailab2025slm}
{NVIDIA Jetson AI Lab}.
\newblock Tutorial - small language models (slm).
\newblock \url{https://www.jetson-ai-lab.com/tutorial_slm.html}, 2025.
\newblock Accessed: 2025-10-01.

\bibitem{tenent}
Zhirui Huang, Rui Ma, Shijie Cao, Ran Shu, Ian Wang, Ting Cao, Chixiao Chen, and Yongqiang Xiong.
\newblock Tenet: An efficient sparsity-aware lut-centric architecture for ternary llm inference on edge.
\newblock {\em arXiv:2509.13765}, 2025.

\end{thebibliography}

\appendix

\end{document}